\documentclass[useAMS,usenatbib,usegraphicx,nofootinbib]{mn2e}
\usepackage{times}
\usepackage{amssymb}
\usepackage{epsfig}

\newcommand{\rtwo}{${R}_{\rmn{200}}$}
\newcommand{\rfive}{${R}_{\rmn{500}}$}
\newcommand{\kmag}{${K}_{\rmn{S}}$}
\newcommand{\kstar}{${K}^{\ast}_{\rmn{S}}$}
\newcommand{\sfryr}{$\rmn{M}_{\odot}~\rmn{yr}^{-1}$}
\newcommand{\msun}{$\rmn{M}_{\odot}$}

\title[BLAST observations of A3112]{Submillimetre observations of galaxy clusters with BLAST: the star-formation activity in Abell 3112}
\author[F. G. Braglia et al.]{ 
  \parbox[t]{\textwidth}{
    Filiberto~G.~Braglia$^{1}$\thanks{e-mail:~fbraglia@phas.ubc.ca},
    Peter~A.~R.~Ade$^{2}$,
	James~J.~Bock$^{3}$,
	Edward~L.~Chapin$^{1}$,
	Mark~J.~Devlin$^{4}$,
	Alastair~Edge$^{13}$,
	Matthew~Griffin$^{2}$,
	Joshua~O.~Gundersen$^{5}$,
	Mark~Halpern$^{1}$,
	Peter~C.~Hargrave$^{2}$,
	David~H.~Hughes$^{6}$,
	Jeff~Klein$^{4}$,
	Gaelen~Marsden$^{1}$,
	Philip~Mauskopf$^{2}$,
	Lorenzo~Moncelsi$^{2}$,
	Calvin~B.~Netterfield$^{8,9}$,
	Henry~Ngo$^{1}$,
	Luca~Olmi$^{10,11}$,
	Enzo~Pascale$^{2}$,
	Guillaume~Patanchon$^{7}$,
	Kevin~A.~Pimbblet$^{14}$,
	Marie~Rex$^{4}$,
	Douglas~Scott$^{1}$,
	Christopher~Semisch$^{4}$,
	Nicholas~Thomas$^{5}$,
	Matthew~D.~P.~Truch$^{4}$,
	Carole~Tucker$^{2}$,
	Gregory~S.~Tucker$^{12}$,
	Elisabetta~Valiante$^{1}$,
	Marco~P.~Viero$^{8}$,
	Donald~V.~Wiebe$^{1}$
	}
  \\
  \\
$^{1}$Department of Physics \& Astronomy, University of
  British Columbia, 6224 Agricultural Road, Vancouver, BC V6T~1Z1,
  Canada\\
$^{2}$School of Physics \& Astronomy, Cardiff University, 5
  The Parade, Cardiff, CF24 3AA, UK\\
$^{3}$Jet Propulsion Laboratory, Pasadena, CA 91109-8099,
  USA\\
$^{4}$Department of Physics \& Astronomy, University of
  Pennsylvania, 209 South 33rd Street, Philadelphia, PA, 19104, USA\\
$^{5}$Department of Physics, University of Miami, 1320
  Campo Sano Drive, Coral Gables, FL 33146, USA\\
$^{6}$Instituto Nacional de Astrof\'isica \'Optica y
  Electr\'onica (INAOE), Aptdo. Postal 51 y 72000 Puebla, Mexico\\
$^{7}$Universit{\'e} Paris Diderot, Laboratoire APC, 10,
  rue Alice Domon et L{\'e}onie Duquet 75205 Paris, France\\
$^{8}$Department of Astronomy \& Astrophysics, University
  of Toronto, 50 St. George Street Toronto, ON M5S~3H4, Canada\\
$^{9}$Department of Physics, University of Toronto, 60
  St. George Street, Toronto, ON M5S~1A7, Canada\\
$^{10}$University of Puerto Rico, Rio Piedras Campus, Physics Dept., Box
23343, UPR station, Puerto Rico 00931\\
$^{11}$INAF, Osservatorio Astrofisico di Arcetri, Largo
  E. Fermi 5, I-50125, Firenze, Italy\\
$^{12}$Department of Physics, Brown University, 182 Hope
  Street, Providence, RI 02912, USA\\
$^{13}$Department of Physics, University of Durham, South Road, Durham, DH13LE, UK\\
$^{14}$Department of Physics, University of Queensland, Brisbane, Queensland, QLD 4072
}

\begin{document}

\date{Accepted ... ; Received ... ; in original form ...}

\pagerange{000--000}

\maketitle

\label{firstpage}

\begin{abstract}
We present observations at 250, 350, and 500 \micron~of the nearby galaxy cluster Abell 3112 ($z$ = 0.075) carried out with BLAST, the Balloon-borne Large Aperture Submillimeter Telescope. Five cluster members are individually detected as bright submillimetre sources. Their far-infrared SEDs and optical colours identify them as normal star-forming galaxies of high mass, with globally evolved stellar populations. They all have $(B-R)$ colours of 1.38 $\pm$ 0.08, transitional between the blue, active population and the red, evolved galaxies that dominate the cluster core. We stack to determine the mean submillimetre emission from all cluster members, which is determined to be 16.6$\pm$2.5, 6.1$\pm$1.9, and 1.5$\pm$1.3 mJy at 250, 350, and 500 \micron, respectively. Stacking analyses of the submillimetre emission of cluster members reveal trends in the mean far-infrared luminosity with respect to cluster-centric radius and \kmag-band magnitude. We find that a large fraction of submillimetre emission comes from the boundary of the inner, virialized region of the cluster, at cluster-centric distances around \rfive. Stacking also shows that the bulk of the submillimetre emission arises in intermediate-mass galaxies ($L < L^{*}$), with \kmag~magnitude $\sim$1 mag fainter than the giant ellipticals. The results and constraints obtained in this work will provide a useful reference for the forthcoming surveys to be conducted on galaxy clusters by {\it Herschel}.
\end{abstract}


\section{Introduction}
\label{intro}

The evolution of cluster galaxies and their star-formation rates have been studied using several different approaches in the last few decades. Optical surveys have shown clear correlations between galaxy colours and local galaxy density or cluster-centric radius (\citealt{Dressler80, Dressler97, Kodama01}). Spectroscopic observations have consistently identified trends in the star-formation activity of cluster galaxies, both as a function of cluster-centric distance (e.g.~\citealt*{Verdugo08}; \citealt{Braglia09} and references therein) and at different redshifts~\citep{Poggianti06, Poggianti09}. Variation of the star-formation rate (SFR) and correlation with the local environment has also been investigated at different mass scales, from groups \citep{Wilman08} to large superclusters \citep{Porter07}, and also in relation to local large-scale filamentary structures (\citealt{Braglia07}; \citealt{Porter08}).

Observations with {\it IRAS} and {\it ISO} provided a way to investigate the nature of dust and to correlate the SFRs of cluster galaxies with their dust content, albeit mostly covering the spectral regions dominated by warm ($> 40$~K) dust. While part of these studies was aimed at detecting diffuse emission from warm intracluster dust (e.g. \citealt{Stickel98}; \citealt{Stickel02}; \citealt{Montier05}; \citealt{Giard08}), several results were also obtained with observations of individual cluster members in several clusters. \citet{Edge01} used combined {\it IRAS}, IRAM and JCMT observations to detect CO line emission from molecular gas in the central galaxies of a sample of 16 cooling core galaxies. Tuffs et al. (2002) and Popescu et al. (2002a; 2002b) observed a large sample of galaxies in Virgo, finding a dependence in the dust content of galaxies with Hubble type. \citet{Pierini03} found that dust luminosity and mass depend on galaxy geometry and shape as well as stellar mass.

Several recent $24\,$\micron~observations with {\it Spitzer}-MIPS have detected dusty star-forming galaxies in intermediate- to high-$z$ clusters. \citet{Geach06} find an increase of the total SFR in clusters with increasing redshift from {\it Spitzer} observations, although with large scatter. \citet{Bai07} investigate the IR properties and the mid-IR luminosity function of cluster galaxies in a higher redshift cluster at $z = $ 0.83, confirming the presence of evolution in the star-formation rate of cluster galaxies. \citet{Fadda08} identify consistent overdensities of $24\,$\micron~sources along two filaments between the clusters Abell 1770 and Abell 1763 ($z = $ 0.23) with respect to the surrounding field. Similar to \citet{Bai07}, \citet{Haines09a} confirm an excess of $24\,$\micron~sources in the cluster Abell 1758 at $z = $ 0.28. \citet{Tran09} also compare the $24\,$\micron~luminous members of a cluster, a supergroup and the field, concluding that the mid-IR inferred SFR is higher in the intermediate environment of the groups than in the field, while it is globally lower in the cluster. Local dependence of the density of 24\,\micron~sources in clusters is investigated in the LoCuSS survey by \citet{Haines09b}, who find a global decrease of star-forming systems with decreasing cluster-centric radius.

Recently, \citet{Wardlow10} have used the AzTEC camera to observe a field centred on the cluster MS0451.6--0305 at $z = 0.54$, identifying two luminous infrared galaxies (LIRG) with a combined SFR of 100 \sfryr. They suggest that, if these are indeed cluster members, they can be examples of a population of galaxies undergoing transformation to the red sequence through interaction with the cluster environment.

All the studies presented have investigated the star-formation activity of cluster galaxies either in the mid-IR or at millimetre wavelengths. However, a complete characterization of the output of star-formation requires coverage of the 200--800\,\micron~spectral region, where the peak of the far-IR emission is expected to lie.

The Balloon-borne Large Aperture Submillimeter Telescope (BLAST: \citealt{Pascale08}; \citealt{Devlin09}) is a pathfinder experiment to {\it Herschel}/SPIRE, and has provided the first maps of selected areas of the sky at 250, 350, and 500\,\micron. These wavelengths were mainly chosen to constrain the peak of the FIR emission from galaxies at redshifts $z \gtrsim 1$. Several studies carried out by the BLAST collaboration on extragalactic fields, either on individual sources (\citealt{Dye09}; \citealt{Dunlop09}; \citealt{Ivison10}), using stacking (\citealt{Devlin09}; \citealt{Marsden09}; \citealt{Pascale09}; \citealt{Ivison10}), or other statistical analyses (\citealt{Patanchon09}; \citealt{Viero09}), have been performed on blank-field maps. A few other studies (\citealt{Rex09}; \citealt{Wiebe09}) have been conducted on known targets. In particular, Rex et al. have provided the first sub-mm maps of the `Bullet' cluster (\citealt{Tucker98}), investigating the nature of a bright sub-mm source identified as a counterpart of a lensed high-$z$ star-forming galaxy. However, no direct investigation of sub-mm emission from cluster members has been conducted so far.

We present here 250, 350, and 500~\micron~observations of a field centred on the nearby cluster Abell 3112 ($z = $ 0.075; A3112 hereafter) carried out by BLAST, and the results of a combined analysis of the optical and sub-mm properties of a spectroscopic sample of its cluster members. These results demonstrate that observation of cluster galaxies at sub-mm wavelengths can provide insight into the star-formation activity in clusters and help understanding galaxy evolution within these overdense environments.

This paper is organized as follows. Section \ref{obsdata} introduces the BLAST observations of A3112 and the ancillary optical and near-IR data used for our study. Section \ref{results} shows the results from stacking analyses of cluster member catalogues and the properties of sub-mm bright cluster members. These results are discussed in Section \ref{discuss} and summarized in Section \ref{summ}.

Throughout the paper, we use a standard $\Lambda$CDM cosmology, where $\Omega_\rmn{M} = 0.3$, $\Omega_{\Lambda} = 0.7$ and $h \equiv H_0/100~\rmn{km}~\rmn{s}^{-1}~\rmn{Mpc}^{-1} = 0.7$.

\section{Observations and data}
\label{obsdata}

\subsection{BLAST}
\label{blastdata}

The BLAST stratospheric telescope has a 1.8-m primary mirror and three broad-band bolometer arrays that observe the sky simultaneously at 250, 350, and 500\,\micron. This array system is effectively a prototype of the SPIRE instrument onboard the {\it Herschel} satellite. The instrument beams are nearly diffraction-limited and are approximately described as Gaussians with full width at half maximum (FWHM) of 36, 42, and 60 arcsec at 250, 350, and 500\,\micron, respectively. For an extended description of the instrument, data analysis, and calibration procedures, see \citet{Pascale08}; \citet{Truch09}.

A large fraction of the successful BLAST observational campaign of 2006 (BLAST06) from McMurdo Station, Antarctica, was dedicated to completing deep and wide blank-field extragalactic surveys (cf. Section \ref{intro}). Smaller fields centred on the positions of well-known targets were also observed (e.g. \citealt{Rex09}; \citealt{Wiebe09}). Among those was a 1.1~$\rmn{deg}^2$ field centred on the nearby cluster A3112 at $z = 0.075$. This target was observed for a total time of 4.2 hr.

The BLAST time-stream data were reduced using a common pipeline to identify spikes, correct detector time drift and calibrate data (\citealt{Pascale08}; \citealt{Truch09}). Maps were generated using the SANEPIC software, which uses a maximum-likelihood algorithm to estimate the optimal solution for the map, as well as producing an associated noise map (\citealt{Patanchon08}). Absolute calibration is based on observations of the evolved star VY CMa and is estimated to be at the level of 10\% (although strongly correlated between the three bands; see \citealt{Truch09} for details).

While the maps represent the optimal weighting of the data across all spatial scales, the largest scales are less constrained due to various systematic effects, particularly because of the lack of cross-linking in the scans of this particular field. This can produce residual large-scale fluctuating patterns across the map. To suppress these spurious signals, all maps have been filtered to remove frequencies contributing to the large-scale noise without affecting the sources, corresponding to scales in excess of about 10 arcmin (approximately the size of the detector array projected on the sky). This procedure, already used for the BLAST GOODS-S (BGS) maps (e.g. \citealt{Devlin09}) also explicitly sets the mean of each map to zero. We obtain $1\sigma$ noise values of 27.3, 21.3 and 15.6 mJy at 250, 350, and 500\,\micron, respectively, in the central 0.8~$\rmn{deg}^2$ of the observed field. This is an area comparable to the BGS-Deep map. An outer, shallower region of about 0.3~$\rmn{deg}^2$ total area was also observed, albeit with higher noise: we calculate $1\sigma$ map r.m.s. of 63.8, 58.6, and 39.0 mJy for this region. Due to the high noise of the wider area, we will only include it in the stacking analyses where it can be appropriately weighted; source extraction will only be performed in the deeper region, to ensure robust identifications.

Analysis of the signal-to-noise maps shows that the observations of A3112 are not strongly dominated by confusion. The distribution of signal-to-noise is well described by a Gaussian with a wing towards high positive values (due to bright sources in the maps), whose width only slightly exceeds pure instrumental noise. We evaluate the contribution of confusion to total measured noise, measured as $\sigma_{\rmn{conf}}/\sigma_{\rmn{map}}$, as 24\%, 26\%, and 29\%  at 250, 350, and 500\,\micron, respectively.

Individual BLAST sources are extracted from each map using a source-finding algorithm which identifies peaks in a smoothed map produced by convolving the flux density map, weighted by the inverse of the variance, with the point spread function (PSF) of BLAST. Peaks with a signal-to-noise ratio (SNR) of at least 3 were selected as sources, their flux then calculated as the value of the beam-convolved flux density map at the position of the peak. The positional uncertainty is calculated as in \citet{Ivison07}, where we assume the slope of the number counts from \citet{Patanchon09} and a minimum uncertainty of 5 arcsec is imposed, equal to the intrinsic pointing uncertainty of the instrument. The positional uncertainty lies below 8, 9, and 13 arcsec for 5$\sigma$ (or more) sources at 250, 350, and 500\,\micron, respectively. Sources at different wavelengths are associated on the basis of their peak position, as in \citet{Devlin09}; the source catalogues are provided in appendix.

We identify 86, 74 and 46 sources at 250, 350, and 500\,\micron, respectively, with SNR values in excess of 3 and up to 23. We use the results of \citet{Patanchon09} to predict the number of sources expected for a blank field with the same area and depth of the A3112 observations. We find expected numbers of $37^{+18}_{-14}$, $22^{+30}_{-16}$, and $20^{+41}_{-17}$ sources at 250, 350, and 500\,\micron, respectively. Despite the relatively large errors on the predicted number counts, the number of sources detected at 250\,\micron~exceeds the expected number by a factor 2. The excess decreases at longer wavelengths, to become completely consistent with the expected counts at 500\,\micron. This suggests that the excess detected is mainly due to the presence of the cluster, whose population should be detected preferentially at 250\,\micron.

\subsubsection{Stacking}
\label{stacking}

Robust association of optical counterparts to BLAST sources is usually possible only for very robust detections and relies on a number of assumptions and on the use of ancillary data (cf. Section \ref{indsources}). Moreover, the map noise will prevent the detection of individual faint sources. It is nevertheless possible to obtain robust statistical information about the average BLAST flux density of a sample of counterparts by {\it stacking} the BLAST maps on a provided catalogue. This technique has been described in great detail by \citet{Marsden09}, and was successfully used to determine the intensity of the far-IR background (FIRB) using BLAST data in the BGS field (\citealt{Devlin09}; \citealt{Marsden09}; \citealt{Pascale09}). Stacking will be applied in Sections \ref{cmembs}, \ref{blaststackrad}, and \ref{blaststackmag}.

\subsection{Ancillary data}
\label{optdata}

BLAST data were combined with optical spectroscopy to identify cluster members based on their dynamical state, and UV to near-IR photometry used to characterize the cluster members on the basis of their photometric properties. This allows us to investigate the FIR star-formation activity of cluster galaxies together with their unobscured UV star-formation and classification from multi-band photometry.

We collected spectroscopic redshifts in the field of A3112 from dedicated observations of the AAOmega Spectrometer on the Anglo-Australian Telescope (AAT). AAOmega \citep{Sharp06} is the new fiber-fed spectrograph for the 2dF robot fibre positioner. It has 392 fibers covering a total area of 2 deg$^2$; a dichroic allow continuous coverage of the spectral region from 3700--8800 \AA~with a resolution $\lambda/\delta\lambda = 1300$. Data were collected on November 24, 2009 (Proposal ID A103, PI: FGB) for a total of 5.1 hours. Target selection was based on available photometry (see below) to select potential cluster members from optical and near-IR colours. 

Two instrumental setups allowed us to collect 683 spectra. Data reduction was performed using the standard AAOmega pipeline, 2dfdr, available on the AAOmega website\footnote{http://www.aao.gov.au/AAO/2df/aaomega/aaomega.html}. Redshifts were extracted using standard IRAF tasks and then checked individually. We obtained redshifts for 578 out of 683 spectra (a success rate of 87\%), with 550 reliable non-stellar spectra.

Additional spectroscopic redshifts in the field were collected from the NASA/IPAC Extragalactic Database (NED). Most of these are from the Las Campanas Redshift Survey (LCRS, \citealt{Shectman96}) and from the  Two Degree Field survey (2dF, \citealt{Colless01}). A total of 188 redshifts were obtained for this field, with a redshift coverage from 0--0.22, from which we identified 90 galaxies not covered by the AAOmega observations for a total of 640 redshifts within 0.8 degrees from the cluster centre.

Optical photometry in the Harris $B$ and $R$ passbands is available from dedicated observations of A3112 carried out at the Las Campanas Observatory Swope Telescope within the Las Campanas/AAT Rich Cluster Survey (LARCS: \citealt{Pimbblet01}; \citealt{Pimbblet02}). These two filters cover the spectral region from 3500 to 6800\,\AA, which includes a number of important spectral features for galaxies at low redshift. In particular, the two filters bracket the 4000\,\AA~break for galaxies at the redshift of A3112, thus providing a classification of galaxies with respect to their global, unobscured star-formation activity by means of the $B-R$ colour. Near-IR photometry was extracted from the 2MASS database \citep{Skrutskie06}, providing $J$, $H$, and \kmag~band photometry for all cluster members. We used the magnitude values from the Extended Source Catalogue (XSC) where available; where the objects were not present in this catalog, the Point Source Catalogue (PSC) was used instead. Undetected galaxies were assigned an upper magnitude limit consistent with the 2MASS limits. {\it GALEX} near- and far-UV photometry was obtained from the Nearby Galaxy Survey (NGS) in the GR4/5 data release. Where no significant UV flux was detected, we assigned an upper limit consistent with the NGS limits in the field of A3112. All optical and near-IR magnitudes were converted to the AB system.

\begin{figure*}
\includegraphics[width=\textwidth]{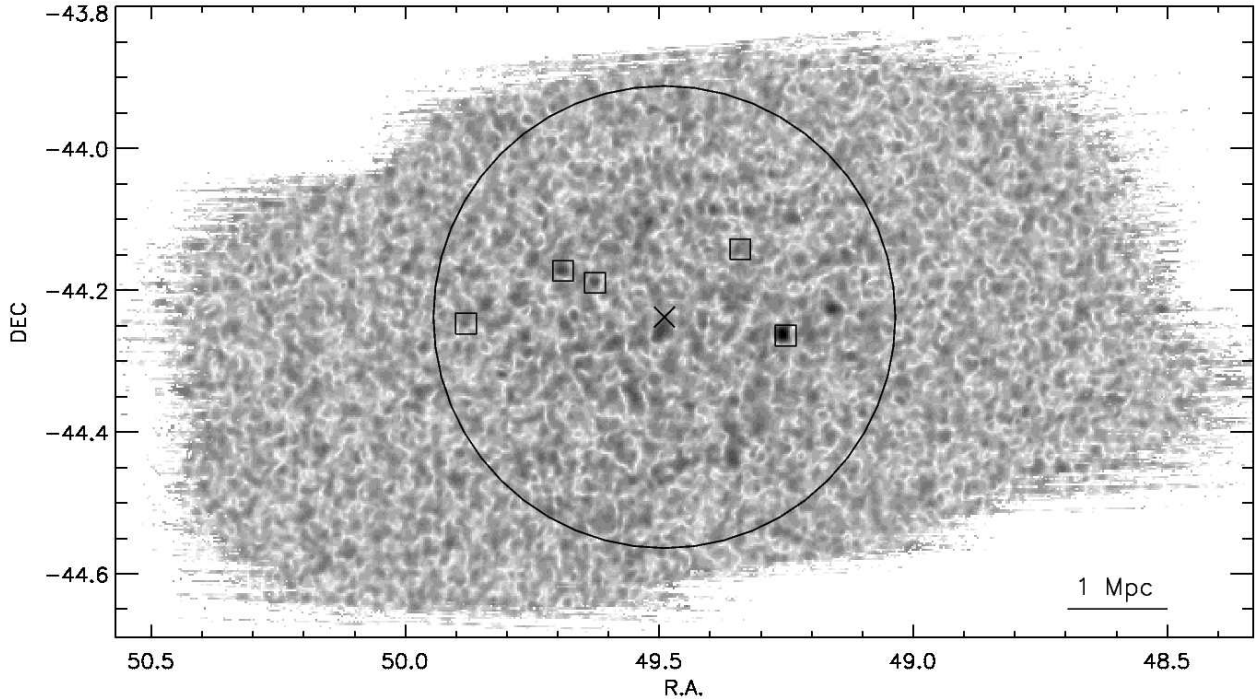}
\caption{BLAST total signal-to-noise map of A3112. The circle marks \rtwo=1.79 Mpc (as calculated from cluster members dynamics, cf. Section \ref{cmembs}). The boxes mark the position of BLAST sources identified as cluster members (Section \ref{indsources}). The cross marks the position of the brightest cluster galaxy and the cluster centre. The distance scale is calculated at the cluster redshift of 0.075.}
\label{clusmap}
\end{figure*}

\section{Results}
\label{results}

\subsection{Cluster members and specific SFR}
\label{cmembs}

Optical cluster members are identified using the technique presented in \citet{Biviano06}. We identify 146 dynamically bound optical cluster members in the field observed with BLAST, scattered across the cluster area from its core to a maximum distance of 3.4 $h^{-1}$ Mpc (at the cluster redshift). We derive a dynamical mass of (4.7 $\pm$ 0.6) $\times$ 10$^{14} M_{\odot}$ within \rtwo = 1.79 $h^{-1}$ Mpc (\rtwo~is the radius where the overdensity of the cluster with respect to the cosmic background density exceeds 200). This value and the cluster X-ray luminosity of 4.2 $\times$ 10$^{37}$ W (1.1 $\times$ 10$^{11}$ L$_{\odot}$) (\citealt{Boehringer04}) identify A3112 as an intermediate mass system.
 
We fit the cluster red sequence (RS) in the ($B-R$) vs. \kmag~plane. This allows for galaxy classification to be carried out independently with respect to stellar mass (traced by the \kmag~magnitude) and to the unobscured star-formation activity (whose signature can be identified in the $B-R$ colour), thus minimizing any residual colour term in the colour-magnitude plane. We fit the RS through a recursive 3$\sigma$-clipping algorithm, which converges to the fit:

\begin{equation}
\rmn{RS}_{(B-R)} = 2.15 - 0.04 \times K_{\rmn{S}} 
\end{equation}

\begin{figure}
\includegraphics[width=8.5cm]{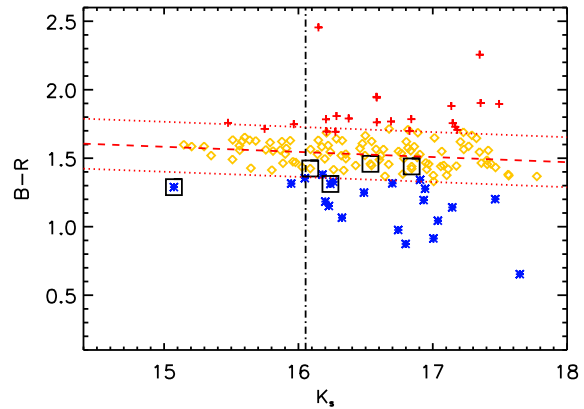}
\caption{Cluster members in the ($B-R$) vs. \kmag~plane. The dashed line shows the fit of the RS with 3$\sigma$ limits shown by dotted lines. Yellow diamonds mark the galaxies along the RS, blue stars the blue star-forming galaxies. Objects redder than the RS are shown as red crosses. The black boxes mark the position of cluster members identified as BLAST counterparts (Sections \ref{indsources}, \ref{blastcmdiag}). The vertical dotted-dashed line marks the value of \kstar, as defined in the text.}
\label{redseq}
\end{figure}

\noindent
Figure \ref{redseq} shows the cluster colour-magnitude diagram. We assume as a reference for \kstar~the value\footnote{\kstar~is the characteristic Schechter magnitude, i.e. the magnitude of the high-luminosity cutoff in the power-law profile of the Schechter luminosity function of the cluster, calculated in the \kmag~band.} obtained by \citet{Cole01} (converted to the AB system) on a large sample of galaxies from 2MASS. 
Out of the 146 identified cluster members, 99 lie along the RS, while 21 of them populate the colour-magnitude plane downwards the RS (see Figure \ref{redseq}). Interestingly, a similar number of cluster members (26) lie upwards the RS. These very red galaxies could in principle be either young and dusty or very old evolved galaxies whose dust reservoir is completely depleted. BLAST observations are a powerful tool to solve this degeneracy without the need for optical spectroscopy, since sub-mm emission is a direct probe of dust.

We derive the unobscured SFR of each cluster member from {\it GALEX} broadband photometry, using the approach described by \citet{Schiminovich05}. On average, cluster galaxies have negligible or very small ($< 1$~\sfryr) SFRs. However, 15 cluster members (i.e. about 9\% of our sample) show moderate SFRs ($\ga 5$~\sfryr).

We also calculate the stellar mass of cluster members from \kmag~photometry using the same recipe provided by \citet{Arnouts07}. We assume the relation for active/blue galaxies (parameter set 1 in Arnouts et al. 2007) for galaxies below the RS, and the relation for quiescent/red galaxies for (parameter set 2) for galaxies along and above the RS. We obtain stellar masses across two orders of magnitude, from $7\times10^{8}$\,\msun~for the smallest galaxies in our sample, to $1.8\times10^{10}$\,\msun~for the central galaxy.

\begin{figure}
\includegraphics[width=8.5cm]{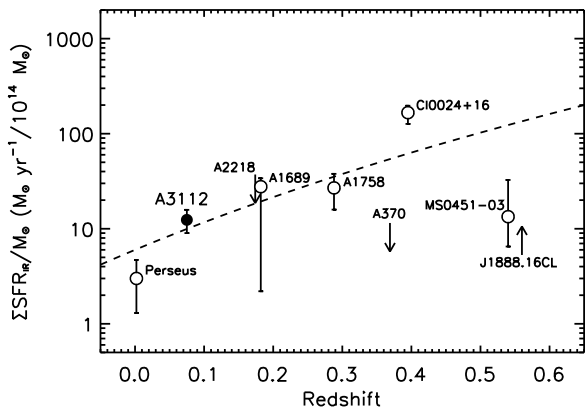}
\caption{The position of A3112 (black circle) in the $\Sigma$SFR vs. $z$ plane. Open circles are data points from \citet{Geach06} and \citet{Haines09b}; the dotted line shows the evolutionary model of \citet{Cowie04}. A3112 agrees closely with the relation found for the evolution of the SFR with redshift.}
\label{specsfr}
\end{figure}

We stack the BLAST flux density maps on the cluster members catalogue to assess the mean flux density in the 3 BLAST bands. We obtain stacked values of 16.6$\pm$2.5, 6.1$\pm$1.9, and 1.5$\pm$1.3 mJy at 250, 350, and 500\,\micron, respectively. This converts to a mean SFR of 0.4~\sfryr~per cluster member and a total SFR of 58.4~\sfryr~for the whole sample of cluster members. We thus calculate a specific SFR, $\Sigma\rmn{SFR} = \rmn{SFR}^{\rmn{tot}}/(\rmn{M}_{\rmn{dyn}}/10^{14} \rmn{M}_{\odot}) = 12.4 \pm 3.4~\rmn{yr}^{-1}$. This is in very good agreement with the relation found by \citet{Cowie04} (Figure \ref{specsfr}; cf. also \citealt{Geach06} and \citealt{Haines09b}).

\subsection{Individual sources}
\label{indsources}

\begin{figure*}
\includegraphics[height=0.31\textheight]{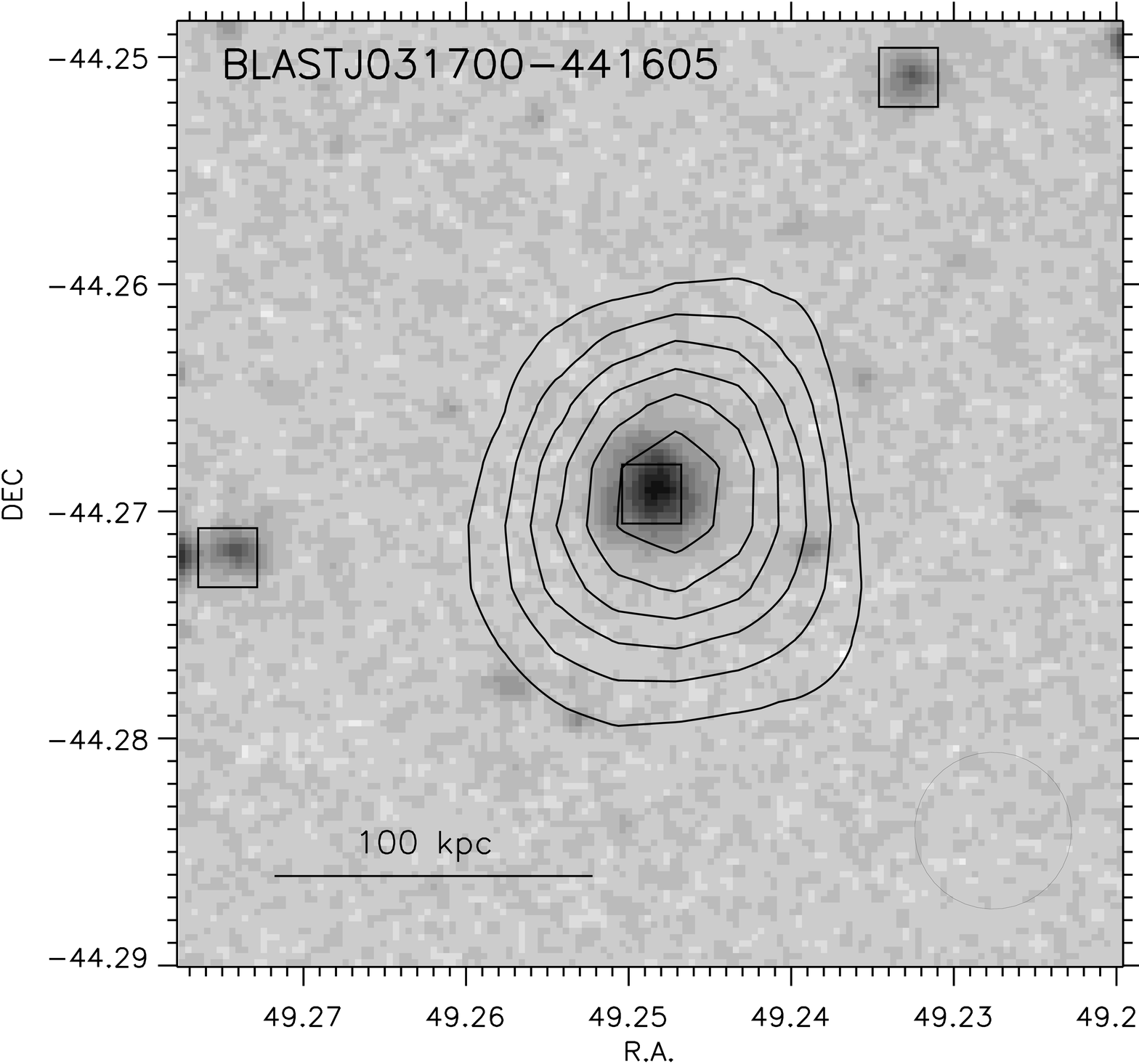}
\includegraphics[height=0.31\textheight]{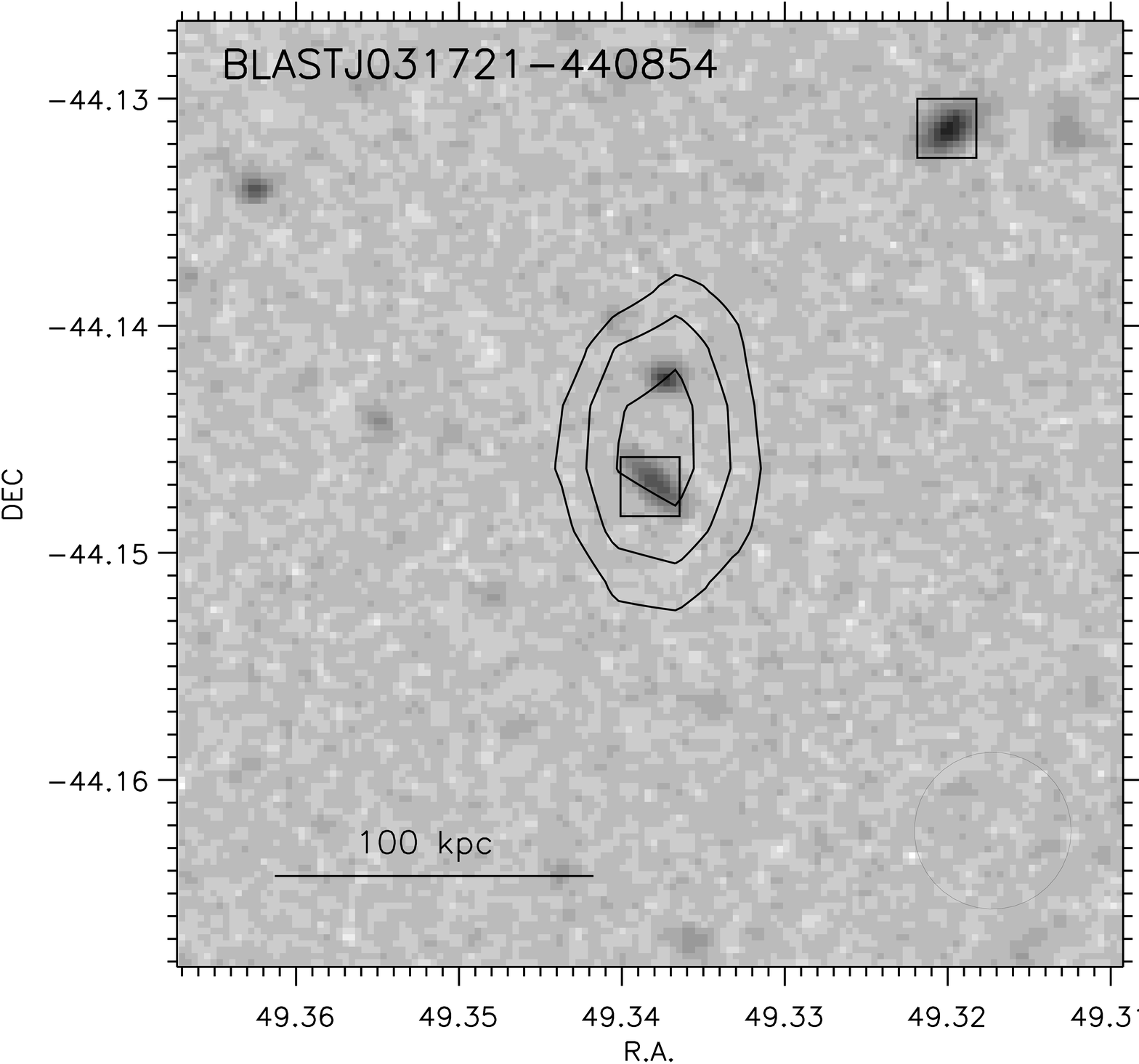} \\
\includegraphics[height=0.31\textheight]{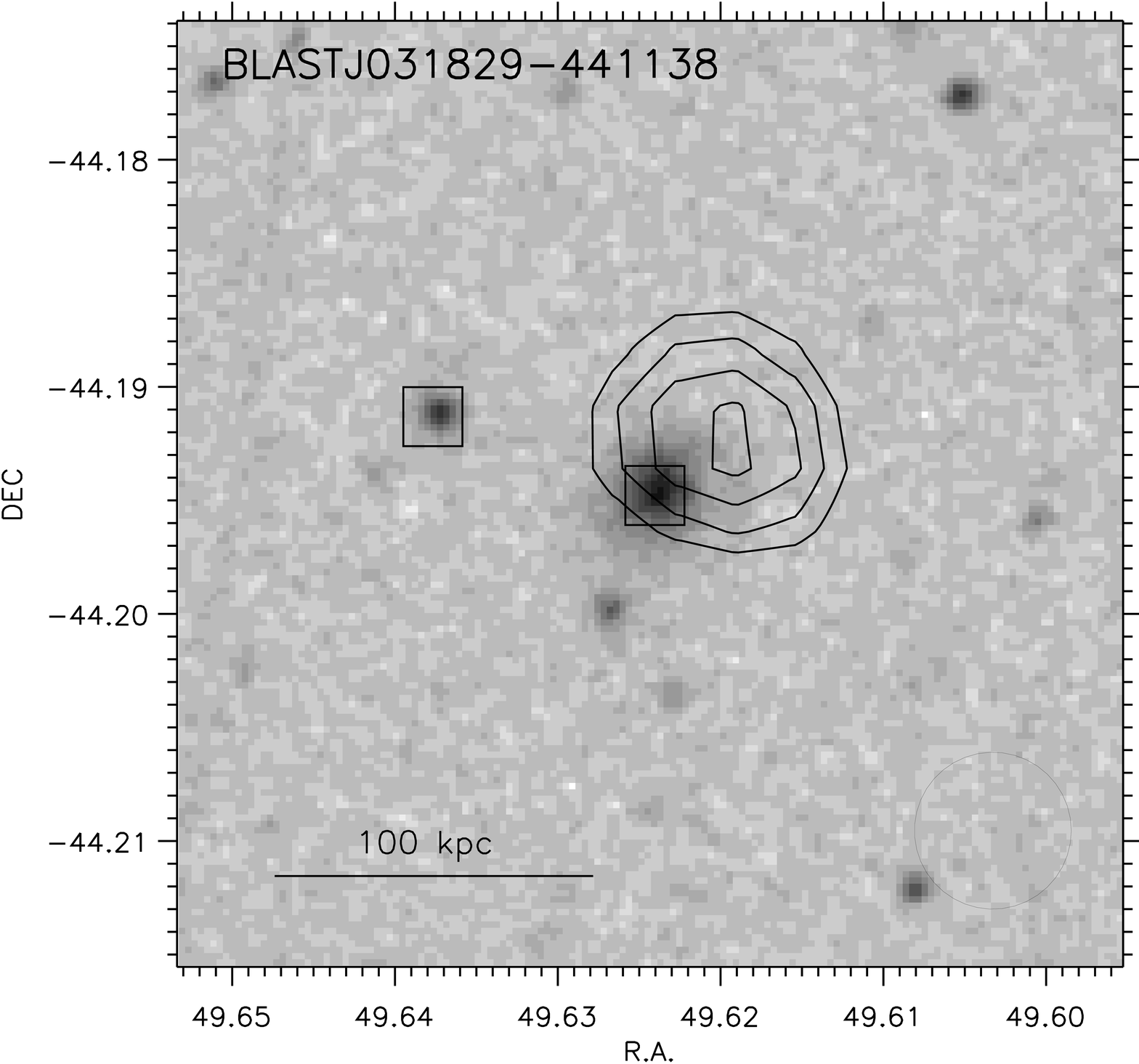}
\includegraphics[height=0.31\textheight]{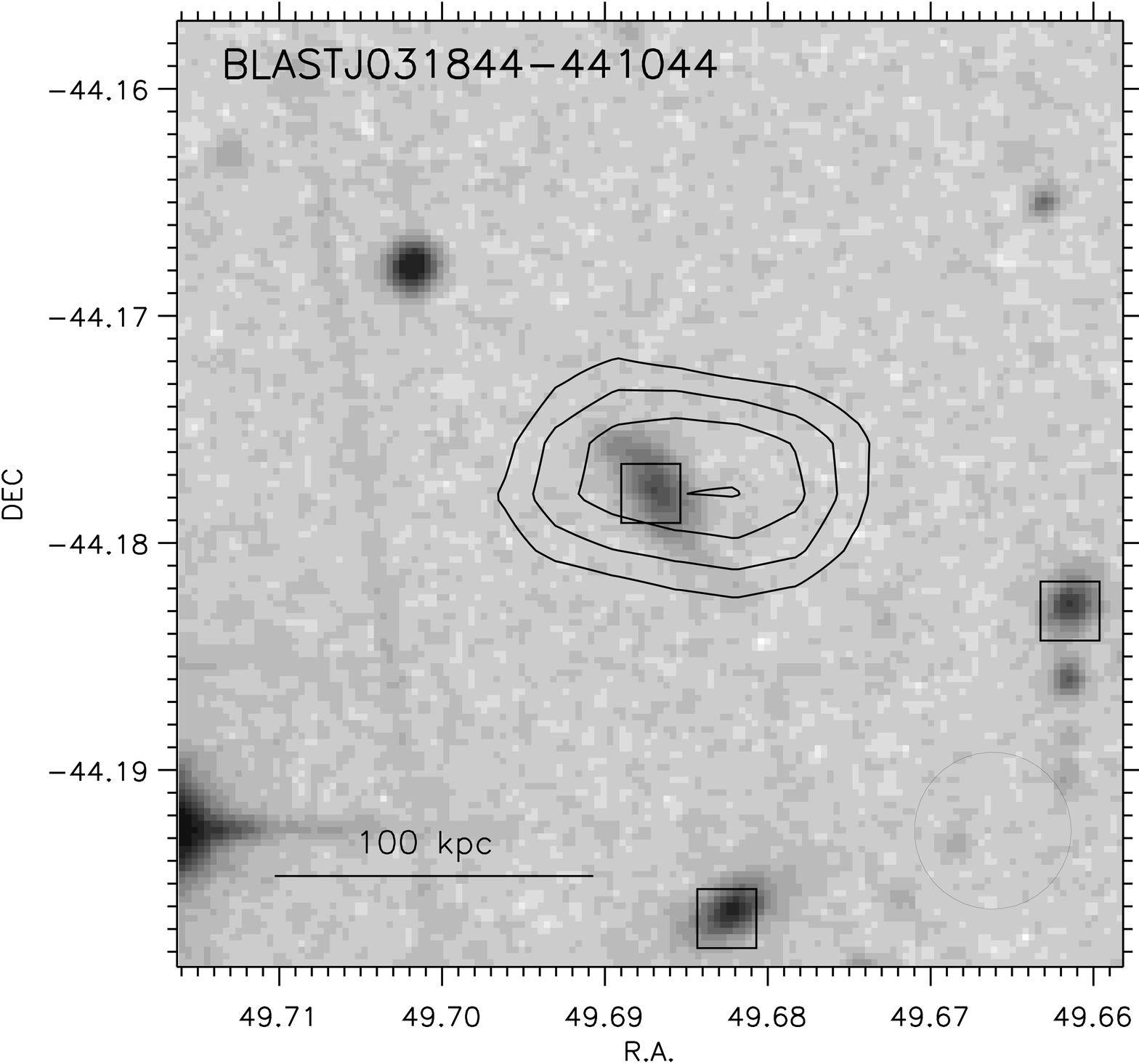} \\
\includegraphics[height=0.31\textheight]{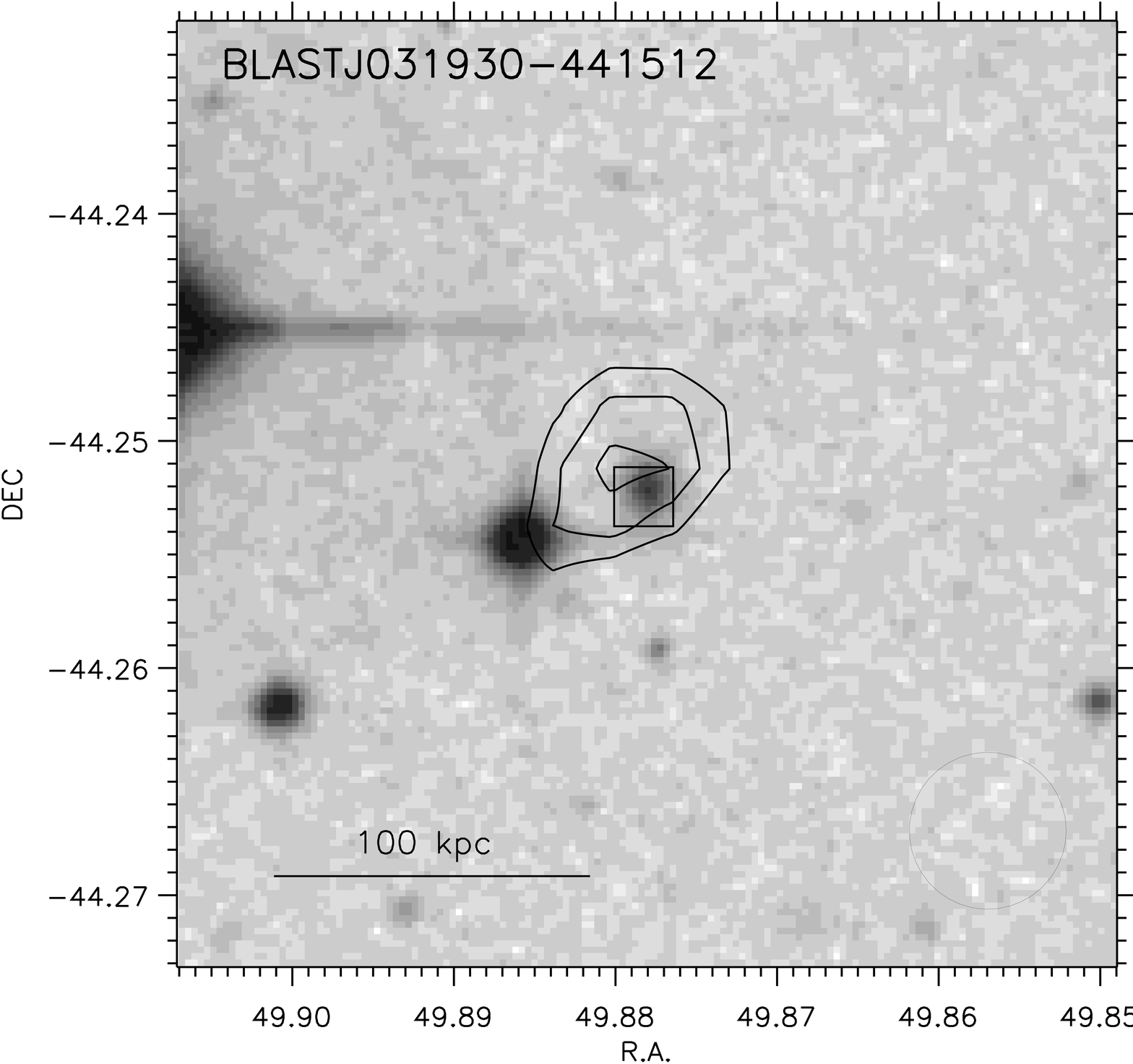}
\caption{The five cluster members identified as counterparts to 250\,\micron~sources. The background images are cutouts from the DSS digitized plates. Boxes mark the position of cluster members, contours show 250\,\micron~emission above a threshold of 3$\sigma$ (~82\,mJy at 250\,\micron). The circle in the lower right corner shows the BLAST 250-\micron~beam size (36 arcsec FWHM). The distance scale is calculated at the cluster redshift. The galaxy on the upper left panel is also an {\it IRAS} source (IRAS~03152--4427).}
\label{cmsources}
\end{figure*}

Far-infrared spectral energy distributions (SEDs hereafter) have been fit to cluster members identified as BLAST counterparts, to help identify their nature. The direct association of optical objects with sub-mm sources is usually far from trivial, mainly due to the large beam size in the sub-mm. Ideally, radio observations would be required to pinpoint the precise position to be matched with optical catalogues (\citealt{Dye09}; \citealt{Ivison10}). However, we notice that in the field of A3112, in each case that a robust ($\geq 5\sigma$) 250\,\micron~source is detected, an optical cluster member is present within 1-$\sigma$ of the positional uncertainty from the source peak. Moreover, these BLAST sources are all found to be brighter at 250\,\micron~than at 350 or 500\,\micron, suggesting that they originate from low-redshift galaxies. As shown later in Section \ref{indsources}, these sources are found to have low temperatures ($T<20$\,K), consistent with being quiescent star-formers at the cluster redshift. Moreover, for these sources a single counterpart is usually detected within the positional uncertainty, ruling out the presence of other low-z galaxies not associated to the cluster that could contribute to the sub-mm signal. We thus conclude that for the scopes of this work, when these criteria are met (i.e. robust detection of a 250-\micron~BLAST source and presence of a single optical counterpart within the positional uncertainty), association of optical cluster members to BLAST sources can be considered robust. In addition to 5$\sigma$ detections at 250\,\micron, we also include sources with a $>4\sigma$ detection both at 250 and 350\,\micron, to increase the number of confirmed counterparts while still maintaining a criterion of robustness. We thus select five galaxies, shown in Figure \ref{cmsources} and whose data appear in Table~1. Of the five selected galaxies, BLAST\,J031700--441605 is also the {\it IRAS} source IRAS~03152--4427. Several other galaxies are in good positional match with BLAST sources, however they are either not robustly detected (i.e.~$< 4\sigma$) at BLAST wavelengths, not cluster members or missing a spectroscopic redshift altogether. We do not include these galaxies in our analysis to avoid biasing our results.

\begin{table*}
\caption{Flux densities (in mJy) at 250, 350, and 500\,\micron~for the cluster members identified as counterparts to 250\,\micron~sources. An asterisk ($^{*}$) denotes a $3\sigma$ upper limit.}
\centering
\begin{tabular}{c c c c c c c c c}
\hline
Source & S$_{100}$ & $\delta$S$_{100}$ & S$_{250}$ & $\delta$S$_{250}$ & S$_{350}$ & $\delta$S$_{350}$ & S$_{500}$ & $\delta$S$_{500}$ \\
\hline
  BLAST J031700--441605 & 960 & 125 & 665 & 29 & 245 & 22 & 79 & 16 \\
  BLAST J031721--440854 & -- & -- & 114 & 28 & 113 & 22 & 64 & 15 \\
  BLAST J031829--441138 & -- & -- & 181 & 28 & 94 & 21 & 54 & 15 \\
  BLAST J031844--441044 & -- & -- & 195 & 28 & 106 & 21 & 63 & 16 \\
  BLAST J031930--441512 & -- & -- & 195 & 34 & 64$^{*}$ & -- & 46$^{*}$ & -- \\
\hline
\end{tabular}
\end{table*}

SEDs are fitted using the same method explained and used in \citet{Chapin08}, taking into account filter response curves, calibration uncertainties and correlations between the BLAST maps (\citealt{Truch09}). Since we have only a modest range of FIR/sub-mm wavelengths (the three BLAST bands), we do not fit the dust emissivity index $\beta$ (i.e. the slope of the modified blackbody); instead we fix it to 2, according to the results of \citet{Wiebe09}. FIR luminosity and dust mass are derived from the SED, and we use the relation by \citet{Bell03} to derive the star-formation rate from the FIR luminosity.

For IRAS~03152--4427, data at 60 and 100\,\micron~are also available from the {\it IRAS} Faint Source catalogue (\citealt{Moshir90}). In particular, 100\,\micron~data allow for better constraints on the shape of the far-IR SED (the 60\,\micron~point already samples the warm dust emission and is not suitable to fit a single-component SED). We find for IRAS~03152--4427 a best-fit temperature $T=(23.4 \pm 0.8)$~K, a total far-infrared (FIR) luminosity of $(1.2 \pm 0.1)\times10^{11}$~L$_{\odot}$, a cold dust mass of $(1.7 \pm 0.1)\times10^8$~M$_{\odot}$ and an SFR of (20.2$\pm$1.9)~\sfryr. This source is thus identified as a luminous infrared galaxy (LIRG) with moderate star-formation. If the SFR derived from the UV (19.4$\pm$0.2~\sfryr, cf. Section \ref{optdata}) is added, we obtain a total SFR of (39.6$\pm$2.0)~\sfryr, still a typical value for LIRGs.

Although other BLAST counterparts lack shorter wavelength data, we use the results from IRAS~03152--4427 as an upper value on their possible luminosity and temperature to help the fitting procedure. We find that the typical cluster members detected at BLAST wavelengths have temperatures of order of 15\,K, FIR luminosities of order of $10^{10}~\rmn{L}_{\odot}$ and dust masses around $10^8$~M$_{\odot}$. Their SFR amounts to a few ($\la 3$) solar masses per year (a few more if the SFR from the UV is also taken into account). Table~2 shows the results of the fit for all selected cluster members, including the derived SFRs and dust mass. Our results show that typical BLAST sources in A3112 are mostly normal star-forming galaxies with low star-formation rates. Fainter sub-mm sources in the cluster are expected to show even lower star-formation activity than the galaxies studied here.

Comparison with previous results (e.g. the {\it ISO} observations of Virgo by \citealt{Popescu02a}) shows that the dust masses are consistent with other observations; our galaxies have dust masses typical of large early-type spirals. This is in agreement with the bright magnitude of these systems, with their optical colours (cf. Section \ref{blastcmdiag}), and with their disc-like optical appearance (Figure \ref{cmsources}).

\begin{table*}
\caption{Result of the fit to the SEDs of the cluster members identified as counterparts to 250\,\micron~sources. The first object is IRAS~03152--4427, whose fit also includes an {\it IRAS} data point at 100\,\micron. Fits for the remaining objects use only BLAST data points. $T$ is in K; $L_{\rmn{FIR}}$ in units of $10^{10}~\rmn{L}_{\odot}$; SFRs in \sfryr; M$_\rmn{D}$ in units of $10^8$~M$_{\odot}$.}
\centering
\begin{tabular}{c c c c c c}
\hline
Source & $T$ & L$_{\rmn{FIR}}$ & SFR$_{\rmn{FIR}}$ & SFR$_{\rmn{FIR+UV}}$ & M$_\rmn{D}$\\
\hline
  BLAST J031700--441605 & 23.4 & 12.0 & 20.2 & 39.7 & 1.7\\
  BLAST J031721--440854 & 11.2 & 1.0 & 1.0 & 2.0 & 7.7\\
  BLAST J031829--441138 & 15.8 & 1.7 & 2.5 & 8.8 & 2.1\\
  BLAST J031844--441044 & 15.1 & 1.6 & 2.3 & 7.0 & 2.7\\
  BLAST J031930--441512 & 17.3 & 1.6 & 2.4 & 9.1 & 1.7\\
\hline
\end{tabular}
\end{table*}

\begin{figure*}
\includegraphics[width=0.49\textwidth]{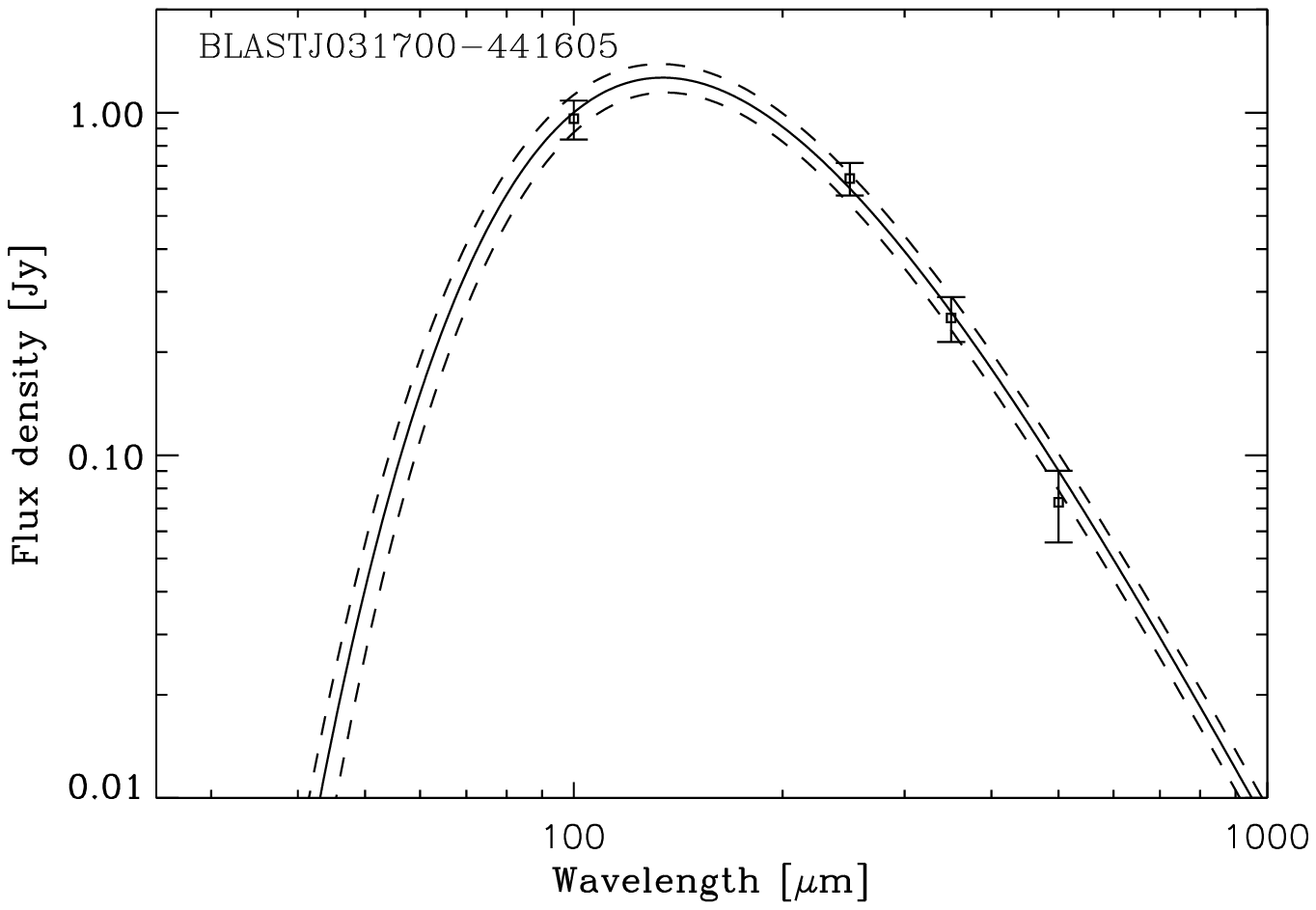}
\includegraphics[width=0.49\textwidth]{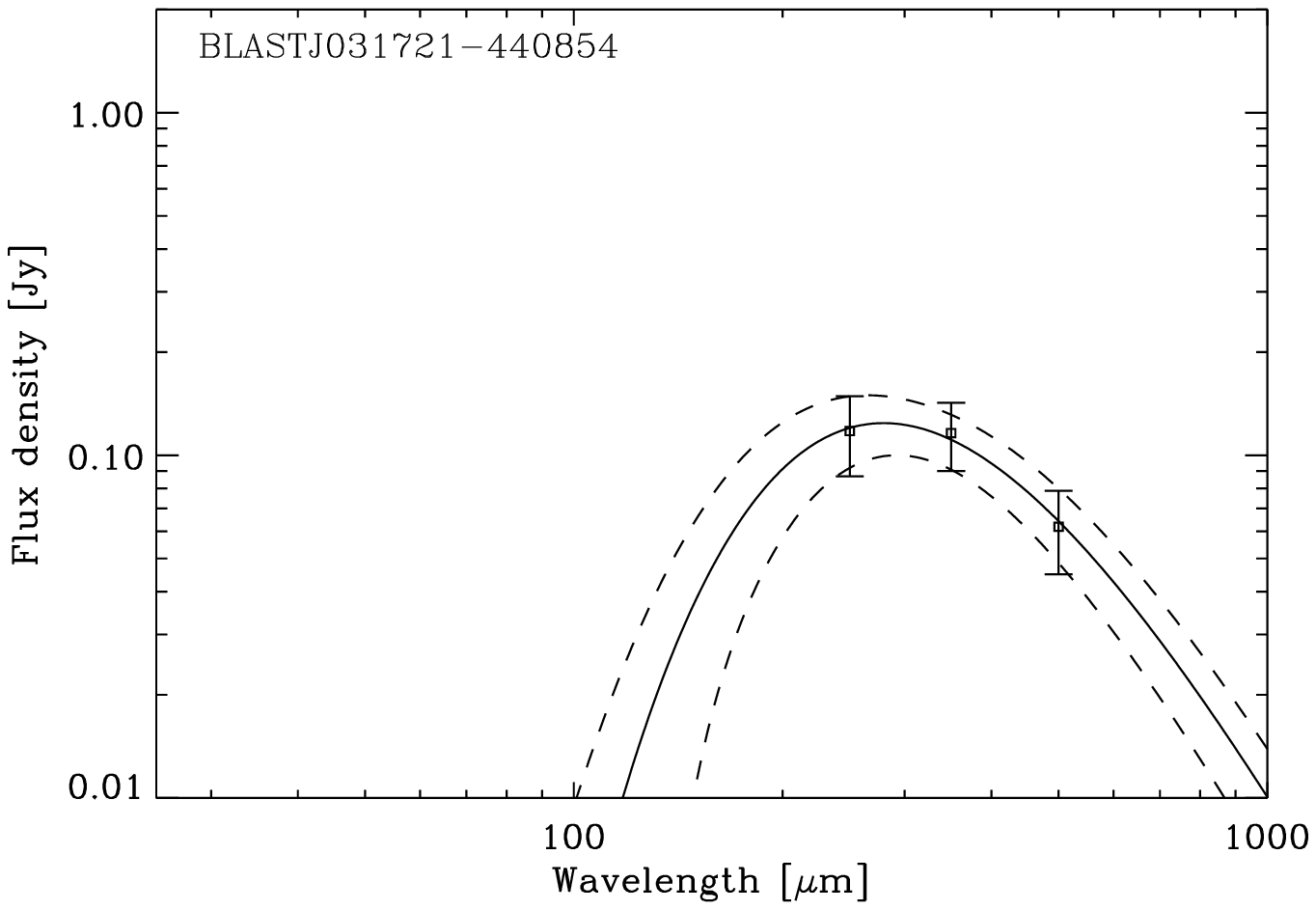}\\
\includegraphics[width=0.49\textwidth]{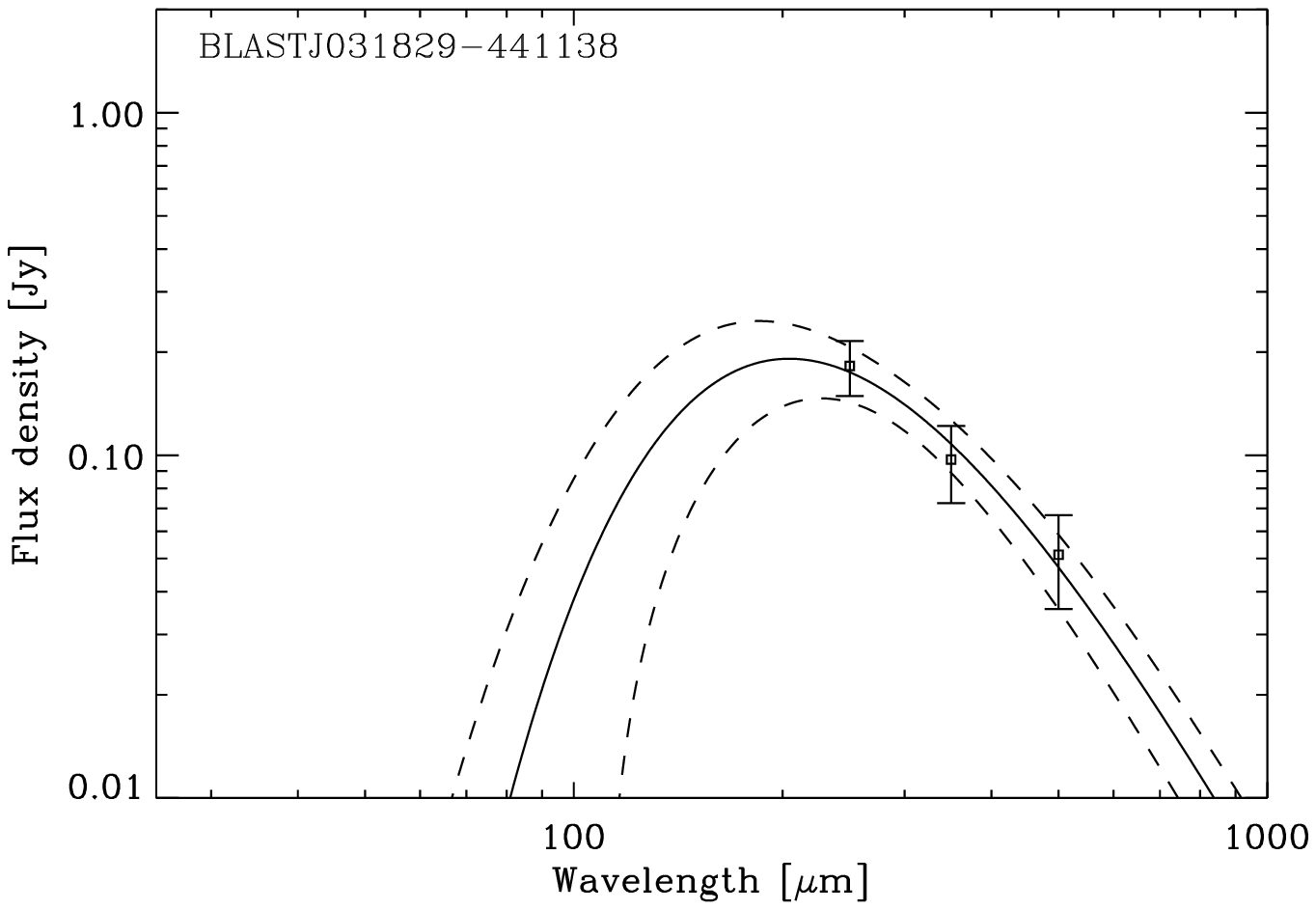}
\includegraphics[width=0.49\textwidth]{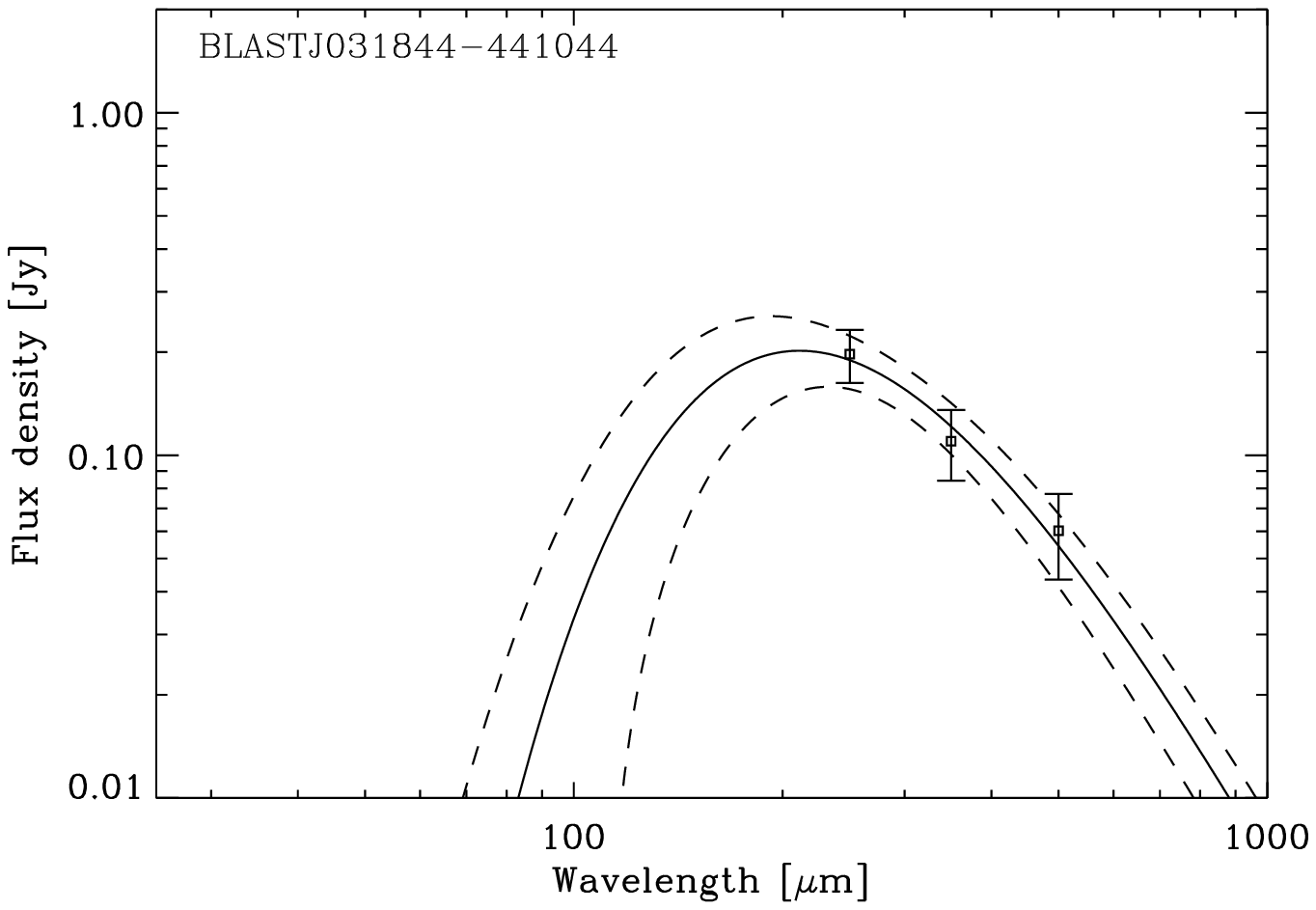}\\
\includegraphics[width=0.49\textwidth]{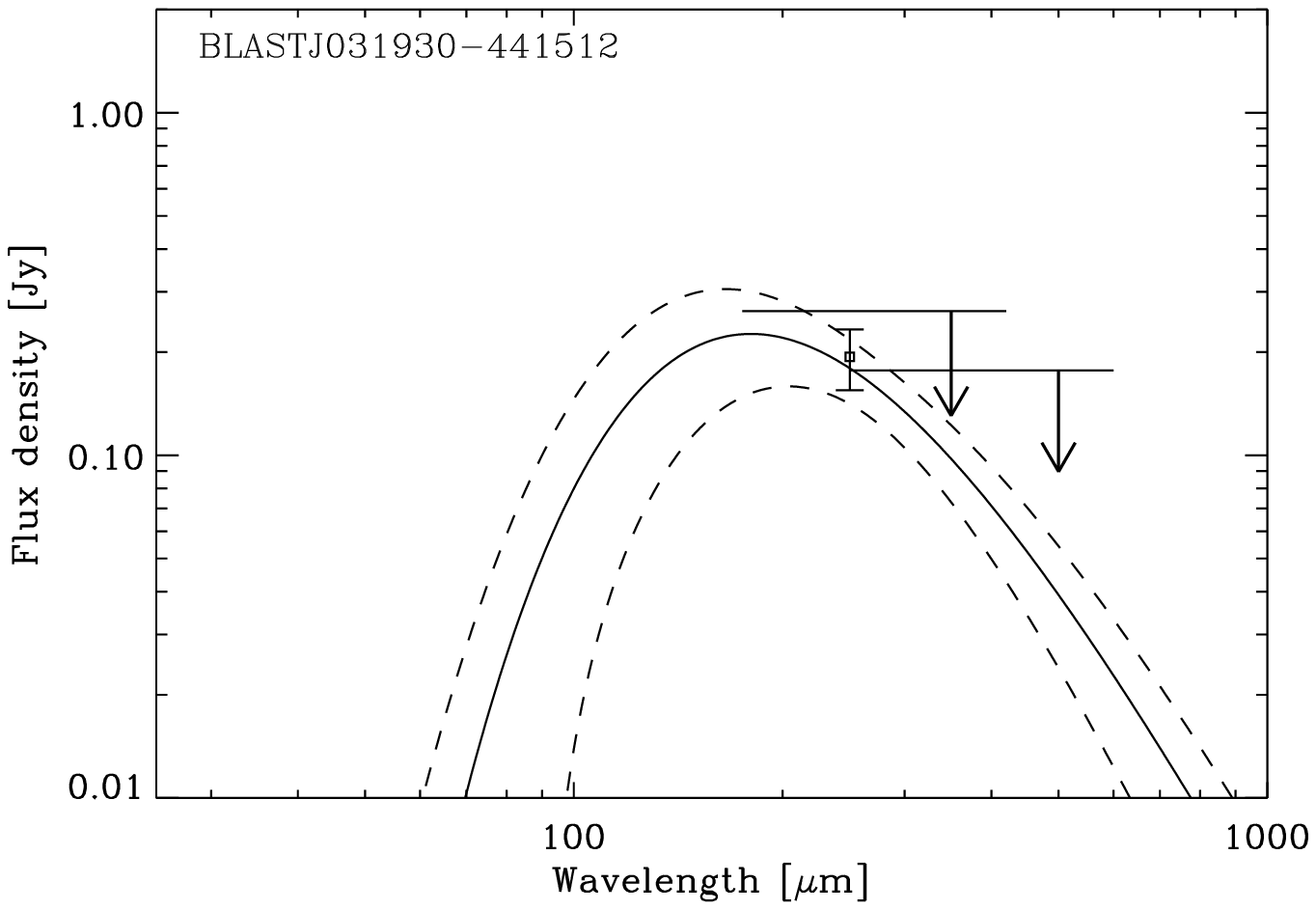}
\caption{The far-IR SEDs of the five cluster members identified as counterparts to 250\,\micron~sources. The solid line shows the best fit to the SED, the dashed lines mark the $1\sigma$ error intervals obtained using a standard Monte Carlo sampling of the parameter space.}
\label{objseds}
\end{figure*}

\subsubsection{Far-infrared emission from the BCG}
\label{bcg}

The brightest cluster galaxy (BCG) in A3112 exhibits strong optical line emission and a powerful radio source. It was targeted with {\it Spitzer} IRAC and MIPS observations by \citet{Quillen08}, who find a weak MIR excess at 24\,\micron. They derive an infrared luminosity of 2.2$\times 10^{10}$~L$_\odot$ which equates to an SFR of 4 \sfryr~(\citealt{O'Dea08}).

The BLAST results are at face value inconsistent with this SFR, with a 2$\sigma$ detection at 250\,\micron~of 57$\pm$28 mJy and upper limits of 64 and 46 mJy at 350 and 500\,\micron~(assuming three times the map r.m.s.). This translates to an upper limit of 3.1$\times 10^{9}$~L$_\odot$ for the FIR luminosity and 0.6~\sfryr~for the SFR. However, the {Spitzer} and BLAST luminosities are derived using different methods and the {\it Spitzer}-derived value is potentially over-estimated due to the MIR contribution of an AGN. The presence of an AGN is consistent with the results of \citet{Takizawa03}, who find an X-ray point source coincident with the core of the BCG and an X-ray excess associated with radio lobes. Spectral analysis identified a power-law component with index 1.9, which confirms this galaxy to host an AGN. The X-ray excess is also explained by \citet{Bonamente07} in terms of relativistic electrons accelerated by a central source. The BLAST limit is at the lower end of the SFR values found both by \citet{{O'Dea08}} and by us with the UV detection of the BCG by GALEX (1.5 \sfryr, cf. Section \ref{cmembs}), which is also likely to be contaminated by the signature of the AGN.

\subsection{BLAST sources along the colour-magnitude diagram}
\label{blastcmdiag}

The cluster colour-magnitude (CM) diagram enables us to select galaxies with respect to their overall unobscured star-formation activity. As already pointed out in Sect. \ref{cmembs}, a large number of objects lie on the cluster RS, identifying them as passively evolving ellipticals. Moreover, we find the region redwards of the RS to be populated by 26 galaxies (i.e., 18\% of the optical sample) showing colours about 0.1--0.2 magnitudes redder than the RS. The combination of BLAST with optical and near-IR data can help in understanding the nature of these systems.

We thus stack the three sub-catalogues of blue galaxies (21 objects below the RS), RS galaxies (99 objects) and very red galaxies (26 objects redder than the RS). We then derive the mean star-formation rate from the FIR luminosity, as in Section \ref{indsources}. The results are presented in Table~3, showing that the majority of the stacked sub-mm emission from our cluster member catalogue comes from optically blue (i.e. star-forming) galaxies. The bulk of their emission is clearly detected at 250\,\micron. On the other hand, RS galaxies show no significant emission in any BLAST band, confirming these systems as largely evolved, dust-free galaxies; consistently, their derived SFR is negligible. The very red galaxies also show little sub-mm emission, mostly consistent with zero, although with large error bars. The inferred SFR is therefore also very low. This suggests that this population is overall depleted of cold dust.

Additional information about the very red galaxies can be inferred from the comparison of their optical and near-IR properties. We find that the ($J$--\kmag) colour of these galaxies is largely consistent with the colour of the RS in the ($J$--\kmag) vs. \kmag~plane, confirming these galaxies to have an old stellar population typical of RS galaxies. Moreover, most of them are found in the central regions of the cluster, where very few young dusty galaxies are expected on average for a non-merging, relaxed cluster like A3112. This suggests that the observed red optical colour could arise from other factors than isotropic dust reddening. 

Several authors in the recent past have shown, using both radiative transfer models (\citealt{Pierini03}; \citealt{Pierini04}) and observational data (e.g. \citealt{Driver07}), that depending on galaxy geometry and orientation, more or less severe reddening can arise. Both \citet{Pierini04} and \citet{Driver07} show that attenuation is largely dependent on observed wavebands, with stronger effects in the bluer bands. A strong contribution is also due to galaxy morphology, the bulge-to-disc ratio playing an important role in determining the overall effect on the observed colours. For intermediate inclination angles ($i < 60$), the ratio of the colour terms $E(B-V)/E(J-K)$ is higher for discs than for bulges. Discs also show overall lower colour terms, regardless of the structure of interstellar dust.

The geometrical parameters derived from SExtractor (namely, the semimajor axis, $a$, and semiminor axis, $b$, of the largest ellipsoidal isophote encircling the galaxy) can be used to roughly estimate the inclination angle of the galaxy along the line of sight (if the observed galaxy is approximated as a disc of radius $R=a$, the inclination angle along the line of sight can be expressed as $\theta_{\rmn{los}} = (90^{\circ}/2\pi)\rmn{acos}(b/a)$). We find that the very red galaxies all have inclination angles between 30$\degr$ and 60$\degr$. This suggests that these galaxies could appear reddened by superposition of a possibly thick disc over a relatively small bulge. In this case, the expected attenuation coefficient is almost negligible in the near-IR, while it can range from 0.2 to about 0.5 in the $B$ band. We thus interpret the very red galaxies as discs with an overall evolved stellar population (i.e. belonging to the RS) and relatively low dust content, which are preferentially reddened because of their inclination. Following this, we stack the sub-mm maps on the combined catalogue of red and RS galaxies, considering them as a single population. We see that the addition of very red galaxies results in a mean SED consistent with that from RS galaxies. Results of stacking are reported in Table~3.

\begin{table*}
\caption{Results of stacking for the three colour-selected catalogs. Fluxes are in mJy; SFR in \sfryr.}
\centering
\begin{tabular}{c c c c c c c c}
\hline
Catalogue & S$_{250}$ & $\delta$S$_{250}$ & S$_{350}$ & $\delta$S$_{350}$ & S$_{500}$ & $\delta$S$_{500}$ & SFR \\
\hline
   Blue & 29 & 6 & 9 & 5 & 4 & 3 & 1.7 \\
   RS & 6 & 3 & 4 & 2 & 2 & 1 & 0.2 \\
   Red & 4 & 5 & 0 & 4 & 6 & 3 & 0.1 \\
   RS + Red & 5 & 3 & 3 & 2 & 2 & 1 & 0.2  \\
\hline
\end{tabular}
\end{table*}

To better characterize the nature of the optical counterparts to the bright BLAST sources, we investigate their location in the CM diagram and their optical colours. Fig.~\ref{redseq} shows the position in the CM diagram of the cluster members identified as BLAST counterparts. These five galaxies all lie in a narrow region of the CM diagram, with a mean colour ($B-R$) = 1.38 $\pm$ 0.08 and \kmag~magnitudes below \kstar~(excluding IRAS~03152--4427, which has a magnitude of order of \kstar).

\subsection{Radial distribution of sub-mm emission}
\label{blaststackrad}

Several studies have revealed the presence of trends in the star-formation activity of cluster galaxies with respect to the cluster-centric radius. This is usually interpreted as a combination of environmental effects that trigger star-formation episodes in galaxies infalling on the cluster from the field. Investigating the radial distribution of sub-mm emission from cluster members can help in better understanding how the environment affects cluster galaxies and to what extent these effects are detected at FIR wavelengths.

We thus divide our spectroscopic catalogue of 146 cluster members into radial bins centered on the BCG, whose position is coincident with the dynamical centre of the cluster. We define the bins so that each contains similar numbers of objects to ensure homogeneous stacking statistics. We find that 25 galaxies per bin is a good compromise between number of bins (6) and S/N ratio.

\begin{figure*}
\includegraphics[width=0.48\textwidth]{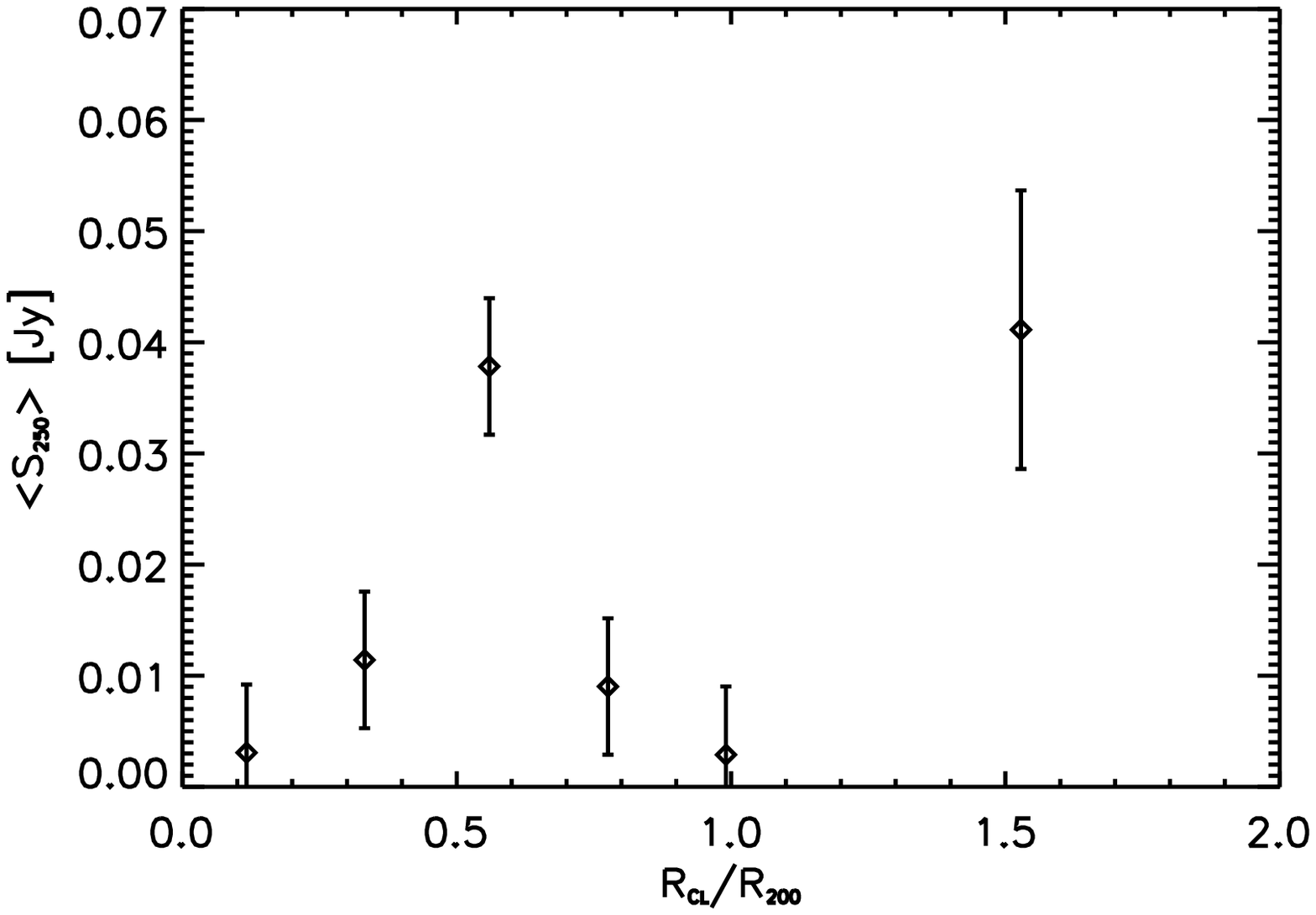}
\includegraphics[width=0.48\textwidth]{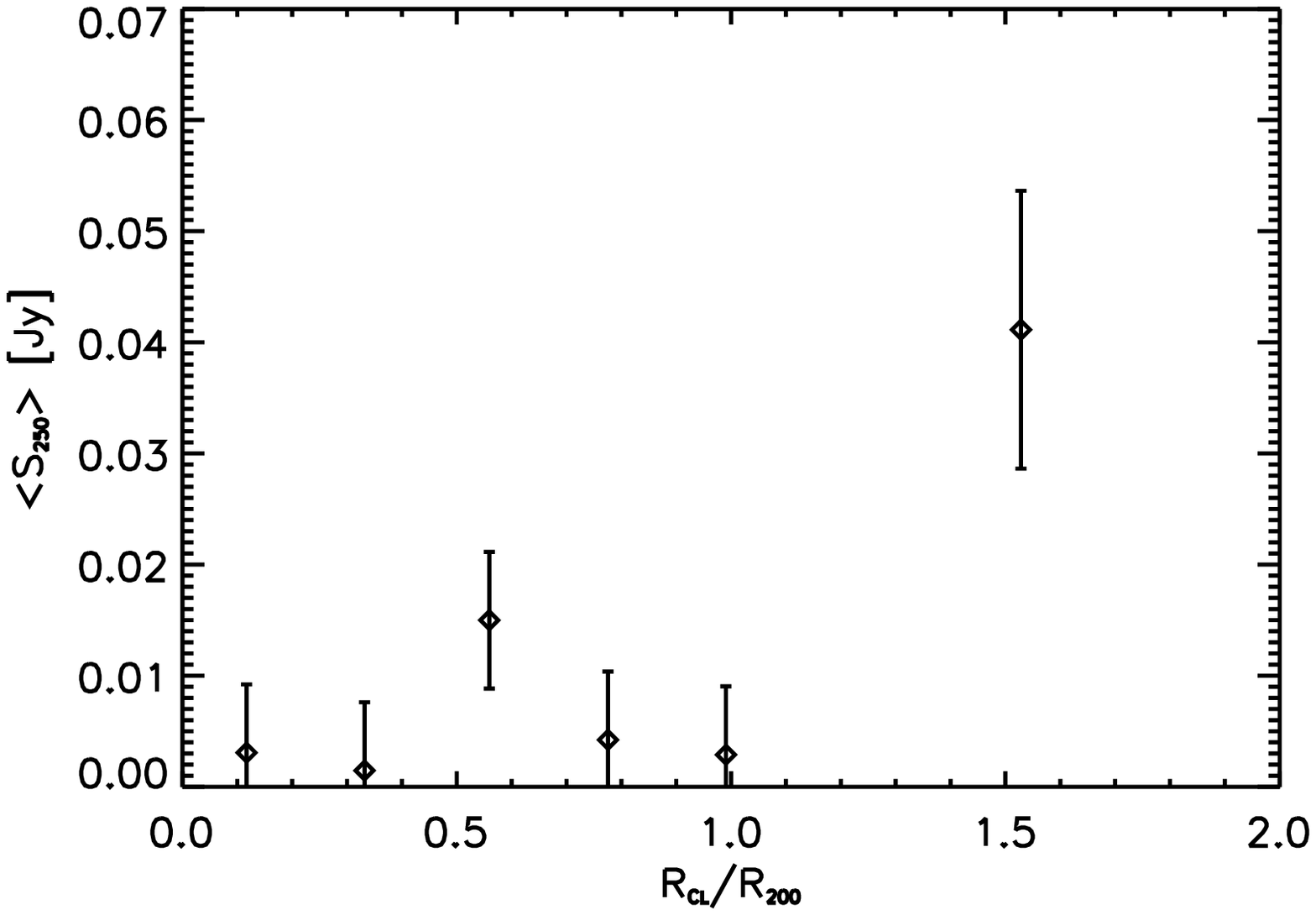} \\
\includegraphics[width=0.48\textwidth]{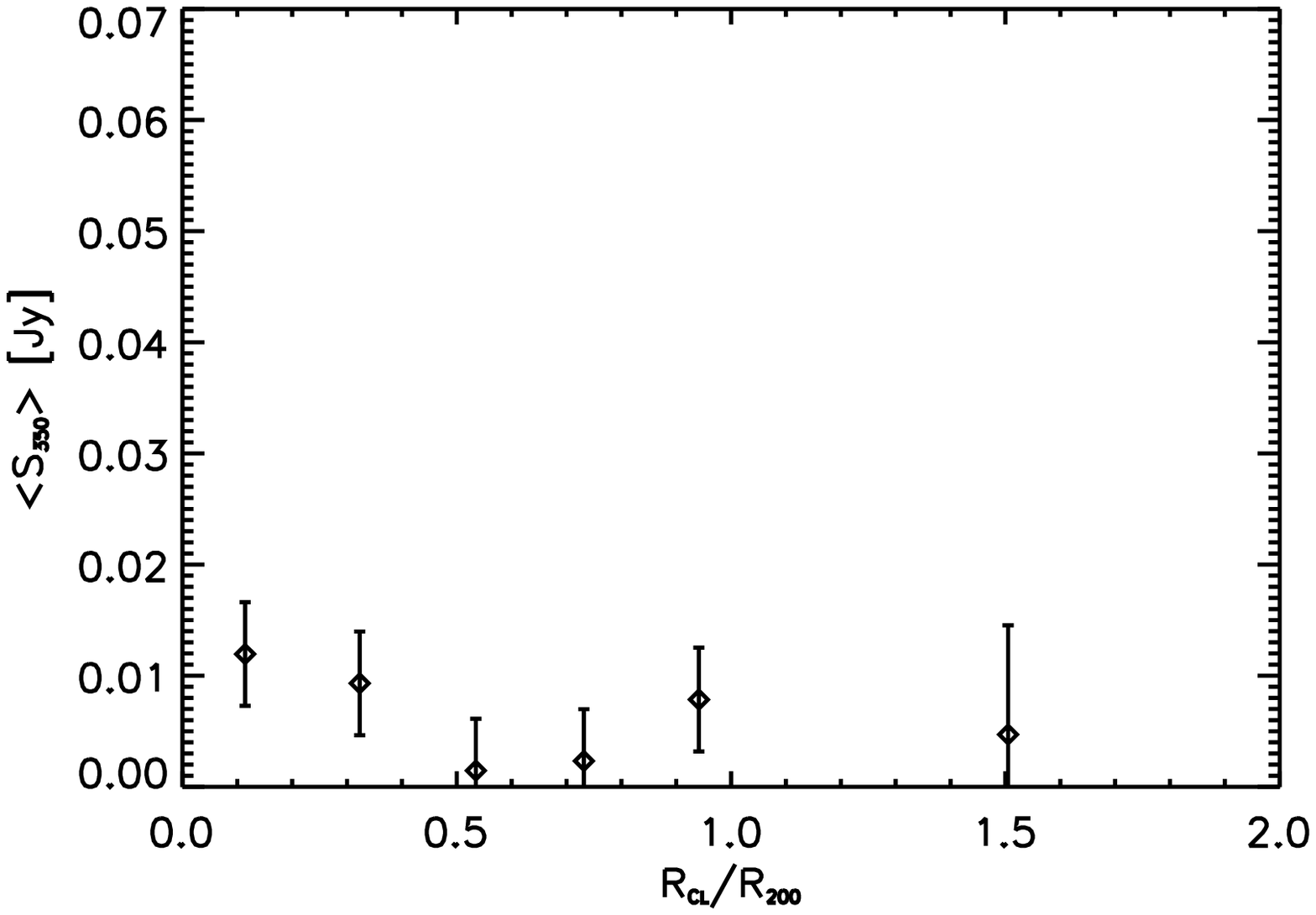}
\includegraphics[width=0.48\textwidth]{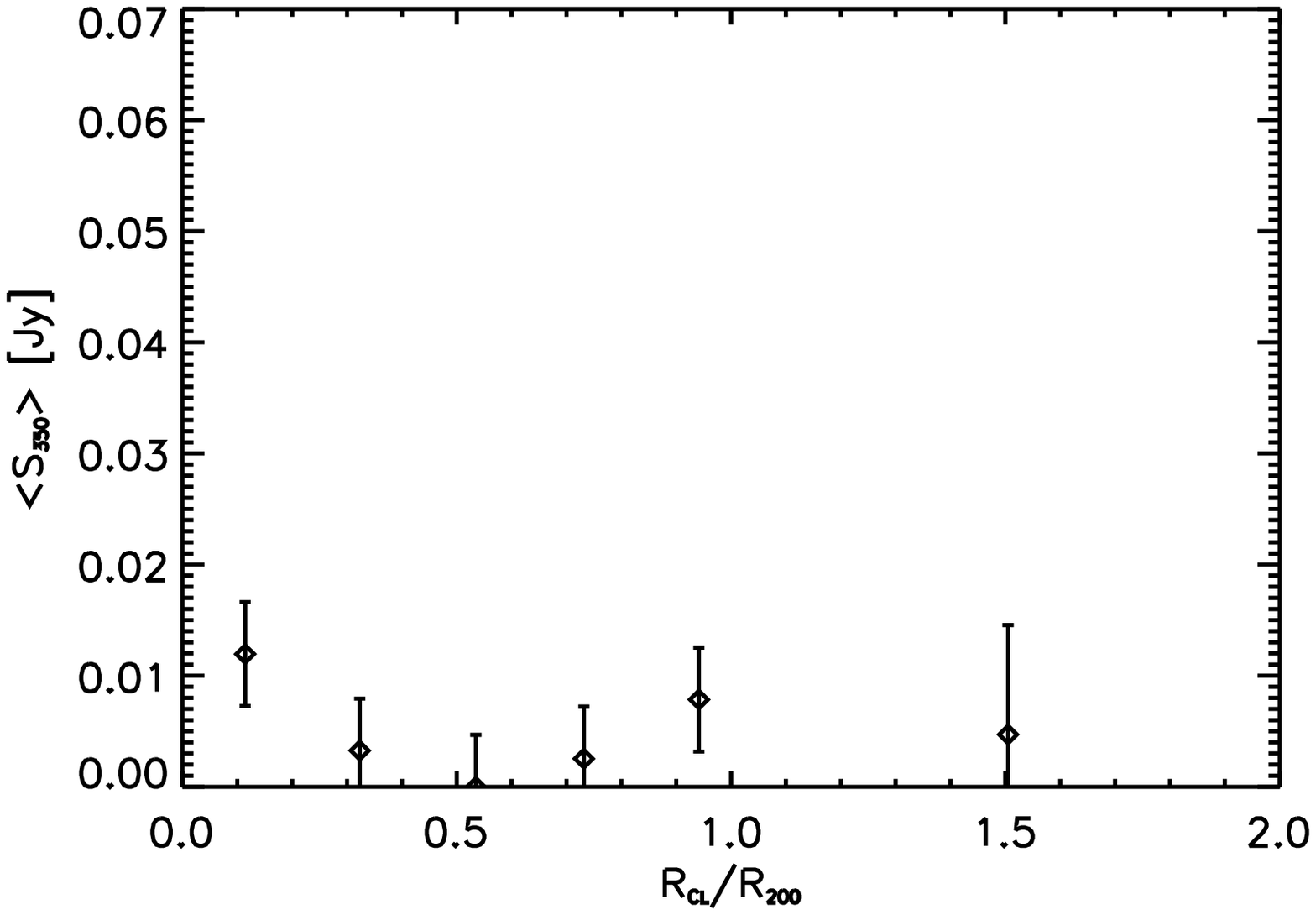} \\
\includegraphics[width=0.48\textwidth]{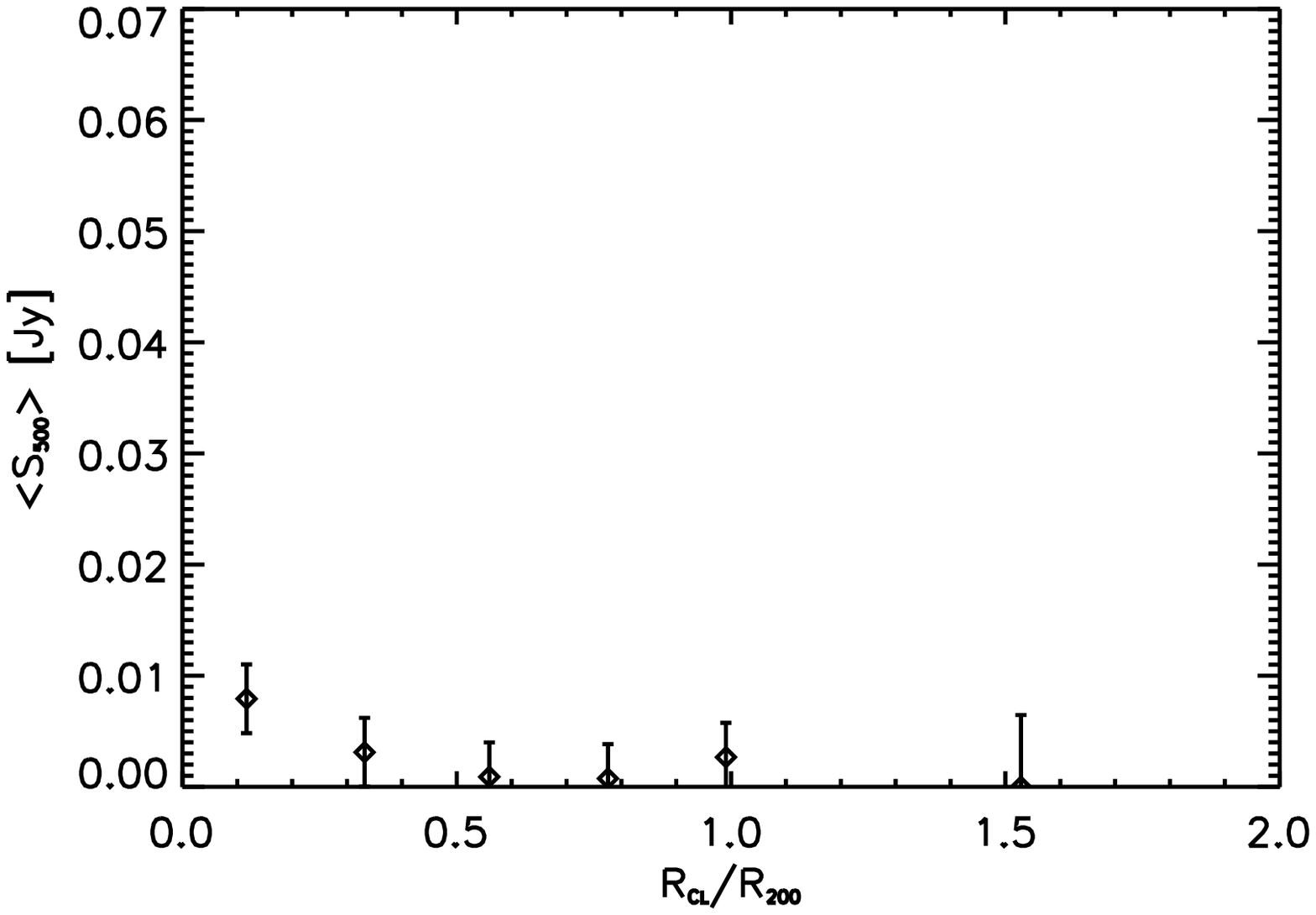}
\includegraphics[width=0.48\textwidth]{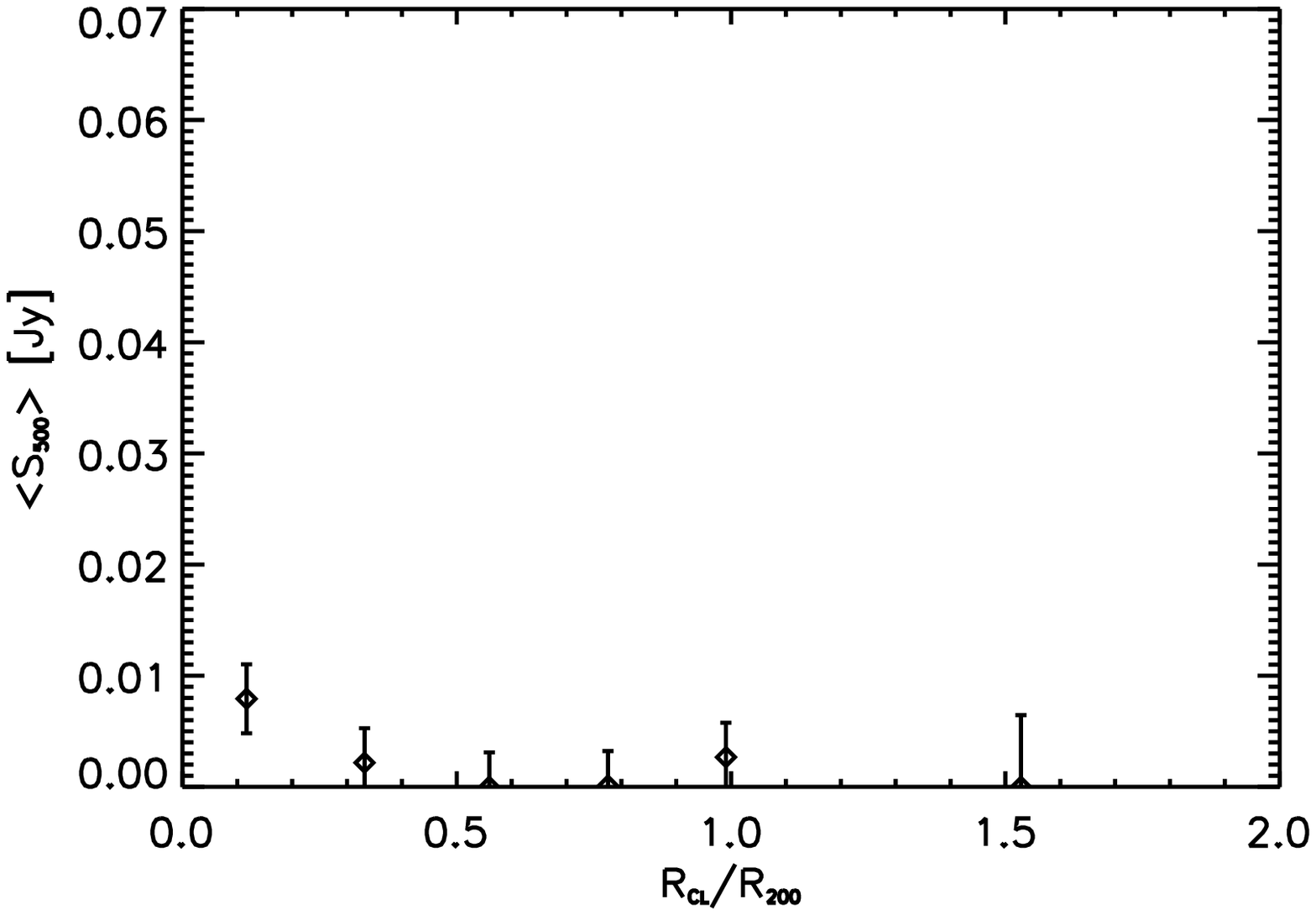} \\
\caption{Stacked BLAST flux of cluster members in bins of different cluster-centric radius ($1\sigma$ error bars). The left panels show the result of stacking the BLAST maps on all cluster members. The right panels show the stacking after removal of cluster members individually detected at sub-mm wavelengths. Top row: stacking of 250\,\micron~map. Middle row: stacking of 350\,\micron~map. Bottom row: stacking of 500\,\micron~map. A significant increase in the mean sub-mm flux density is detected around 0.6\rtwo~at 250\,\micron, along with another increase beyond~\rtwo.}
\label{stackrad}
\end{figure*}

The radial plot of the mean stacked flux density (see Fig. \ref{stackrad}) shows two significant ($\sim 4\sigma$) peaks at 250\,\micron. The outermost of these peaks is detected at about 1.5\rtwo, i.e. outside the gravitational radius of the cluster (this region is only covered by the shallower part of the BLAST map and thus the stacking results are quite noisy). This is expected, since at radii $\gtrsim$\rtwo~the infall of galaxies from the field is expected to be most noticeable, and thus an increase in the mean SFR is expected (e.g. \citealt{Balogh99}). This signal drops at longer wavelengths.

A second peak is detected closer to the cluster core, around 0.6\rtwo. To investigate whether this peak is genuinely due to the presence of the cluster, we use the same method and stack the BLAST-BGS map (Devlin et al. 2009) on simulated catalogues. Since the BGS is a blank field, we can use it to assess the probability of random occurrence of a radial peak with the same significance as detected in A3112. We thus generate 10000 sets of 146 positions, randomly distributed around a centre and on an area equivalent to the map of A3112, stack the BGS map on each, look for the presence of a radial peak and calculate its deviation from the mean of the radial profile. To test if the presence of a cluster affects the outcome of the stacking analysis, we also simulate 10000 catalogues of 146 normally distributed positions over the same area (we use the value of \rfive~found for A3112 as the standard deviation of the distribution, to roughly reproduce a density profile consistent with the presence of a cluster). We find that the probability of random occurrence of a peak as significant as that detected in A3112 is less than $10^{-5}$ in all cases. Limiting this analysis to the BGS-Deep or BGS-Wide maps also yields similar results.

To check independently the significance of this peak, we also compare the stacked values with two radial profiles: a flat profile with value equal to the mean of the stacked values and a Gaussian-shaped profile which we assume as a representation of the peak. A Kolmogorov--Smirnov (KS) test shows that the flat profile is ruled out with a rejection probability of 0.9996, while a peaked profile is accepted with very high probability (rejection of $9\times10^{-4}$). We also test the same profiles by stacking in smaller radial bins (12--20 objects per bin); although the data are noisier due to lower statistics, the flat profile is still ruled out while the peaked profile provides a good fit.

We also notice that the cluster members matching the position of 250\,\micron-detected sources mostly lie at the distance of the inner radial peak. These galaxies do not appear to be clustered among themselves, being instead randomly distributed azimuthally. In order to evaluate whether the peak is entirely due to the presence of these galaxies, we stack our catalogue after removing them. Although the peak shows a lower deviation from the mean value ($\sim 2\sigma$), a flat profile again has a low probability of being a good representation of the data (although with a lower rejection probability of 0.3). A peaked profile has again a very low rejection probability ($1.3\times10^{-3}$), thus providing a robust representation of the data.

The same analysis on the 350 and 500\,\micron~data shows no significant peak at the same position, although cluster members from the innermost bin show on average a non-zero mean flux density. The significance of this signal is nevertheless low ($\sim 1.5\sigma$).

To ensure that no selection effect in the input catalogue is biasing this result, we also investigate the $(B-R)$ colour of galaxies in each bin. We find that all radial bins (but the innermost) are populated by a mix of red-sequence and blue galaxies, thus assuring that the peak is not due to undersampling of blue (i.e. star-forming) galaxies at radii of the order of \rtwo. The number of blue galaxies is indeed the same (4--5) in all radial bins.

\subsection{Sub-mm emission across the luminosity function}
\label{blaststackmag}

We can also stack maps on the spectroscopic catalogue of cluster members after dividing into bins of absolute \kmag~magnitude, to study how the sub-mm BLAST flux is distributed across the cluster luminosity/mass function. Although the available data do not allow us to probe the dwarf population of cluster members, they cover three full magnitudes in the \kmag~band, enough to reach intermediate-luminosity galaxies down to \kstar+2. As in the case of radial stacking, we define magnitude bins containing a constant number of objects (25) for homogeneous stacking statistics.

\begin{figure*}
\includegraphics[width=0.48\textwidth]{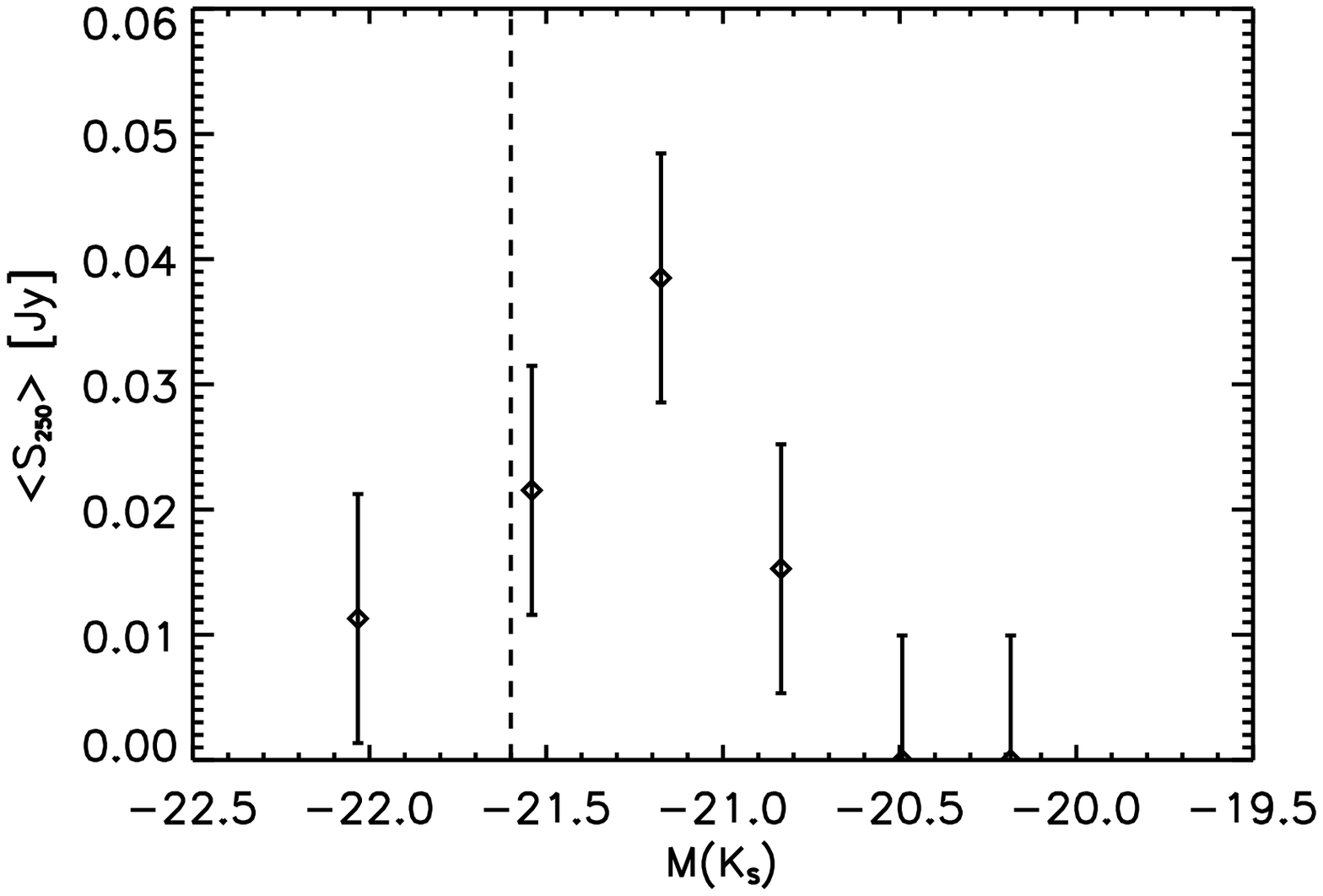}
\includegraphics[width=0.48\textwidth]{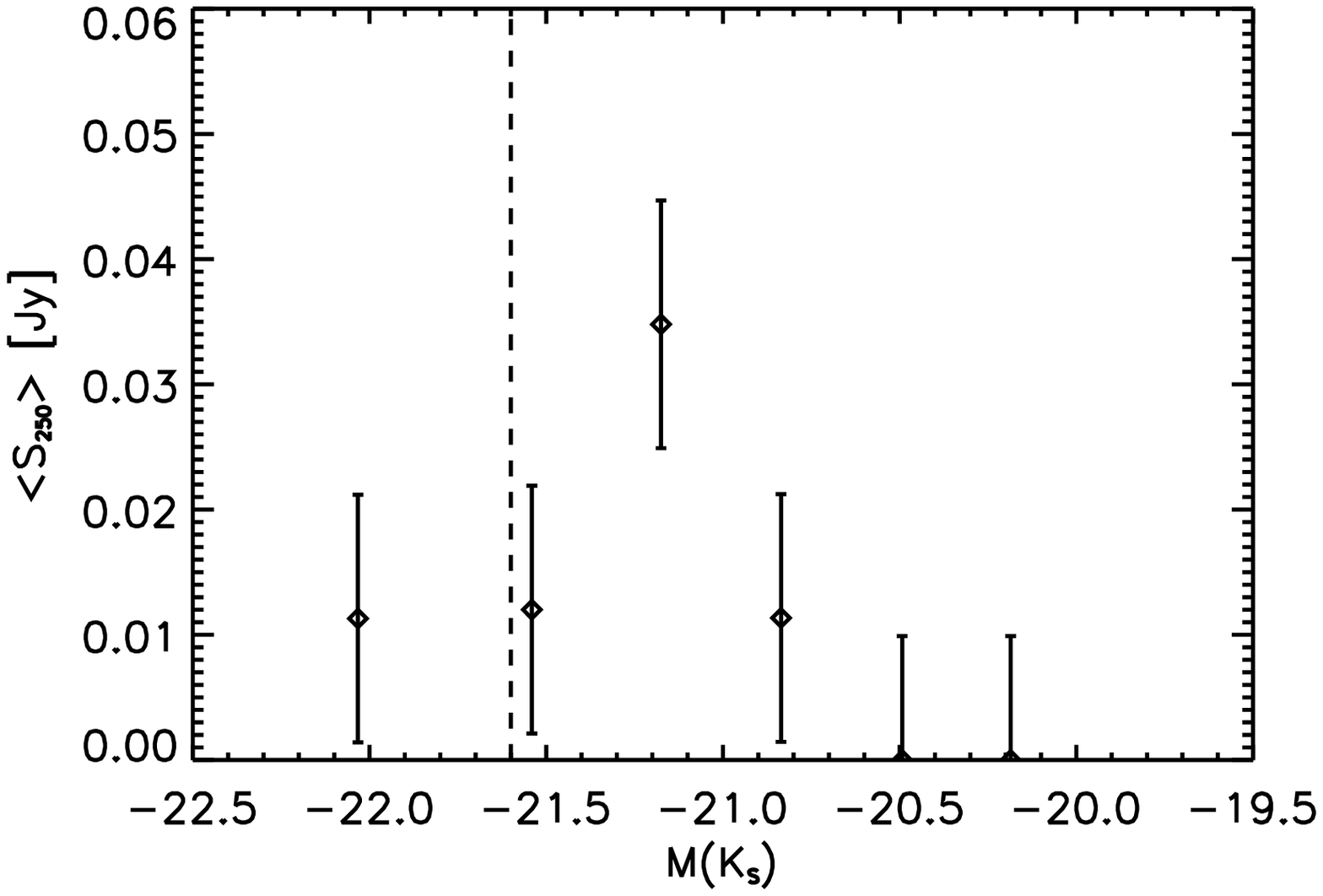}
\includegraphics[width=0.48\textwidth]{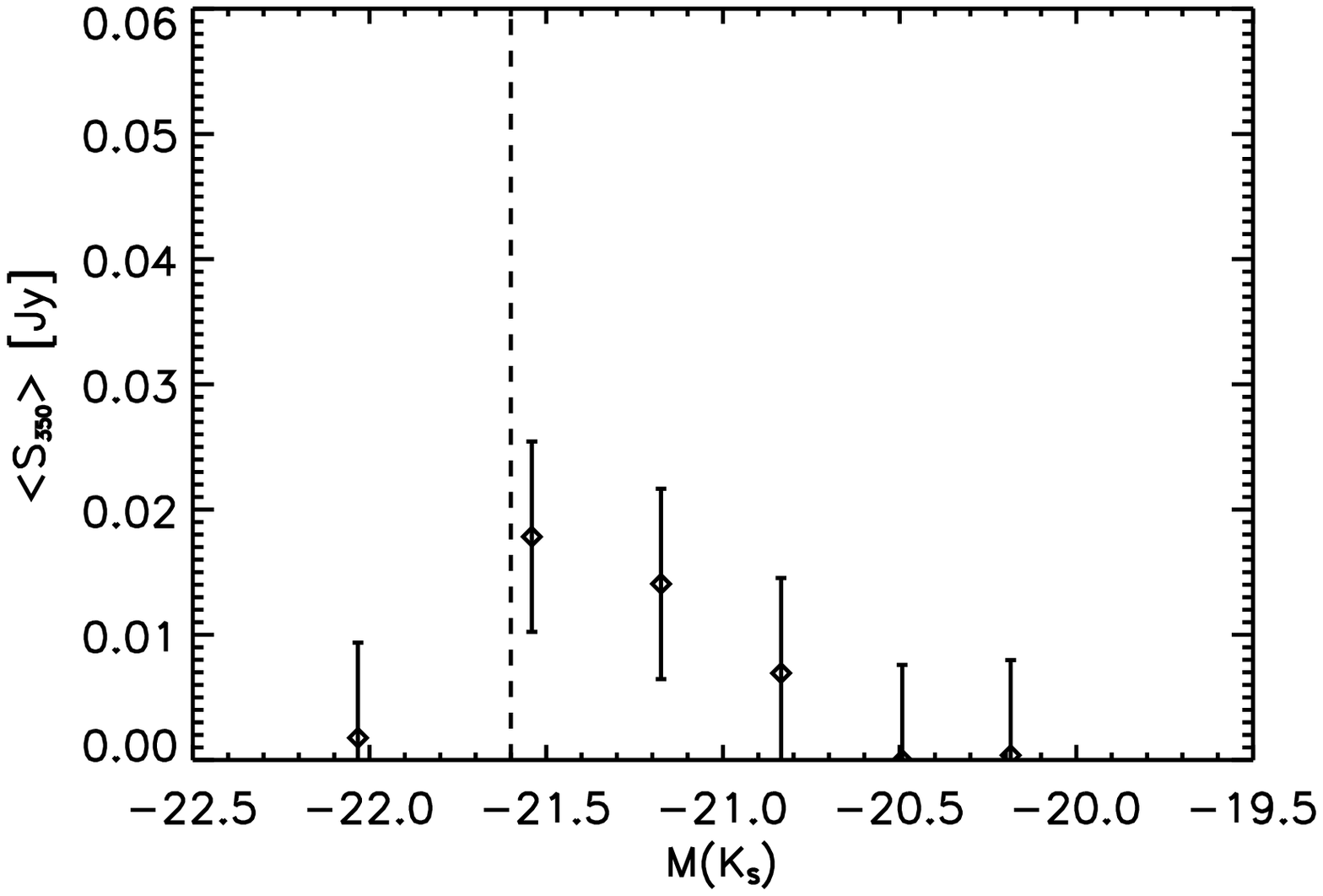}
\includegraphics[width=0.48\textwidth]{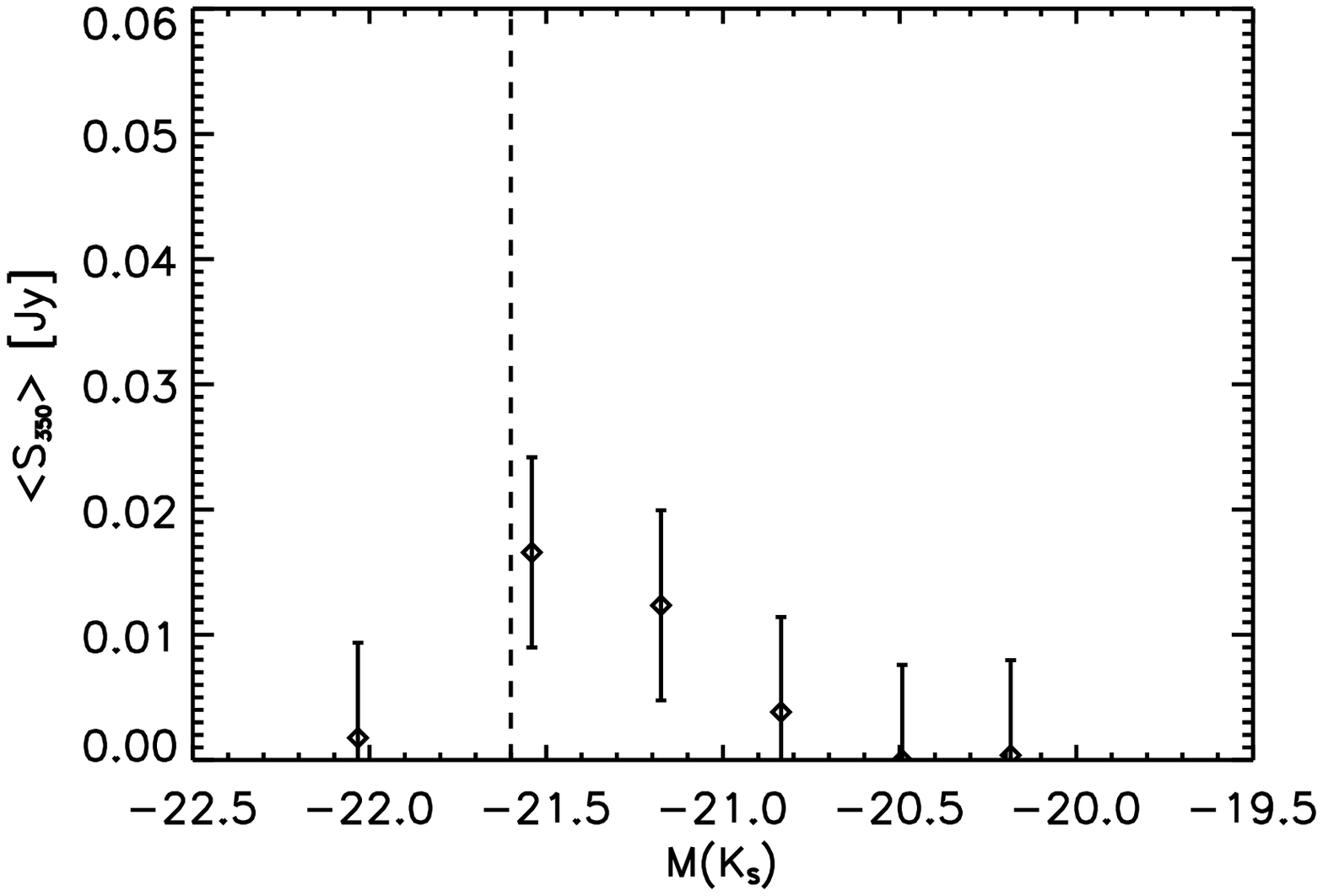}
\includegraphics[width=0.48\textwidth]{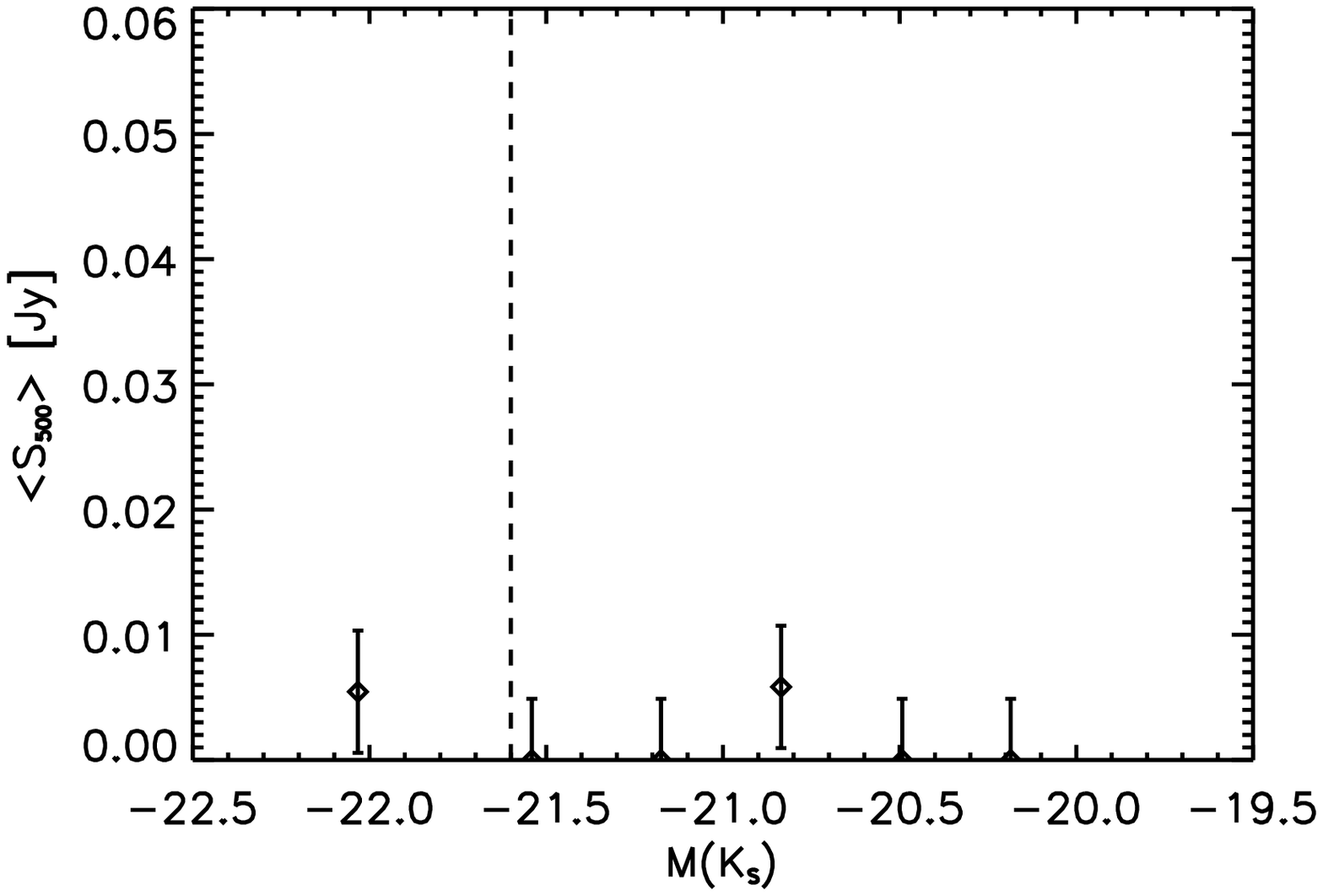}
\includegraphics[width=0.48\textwidth]{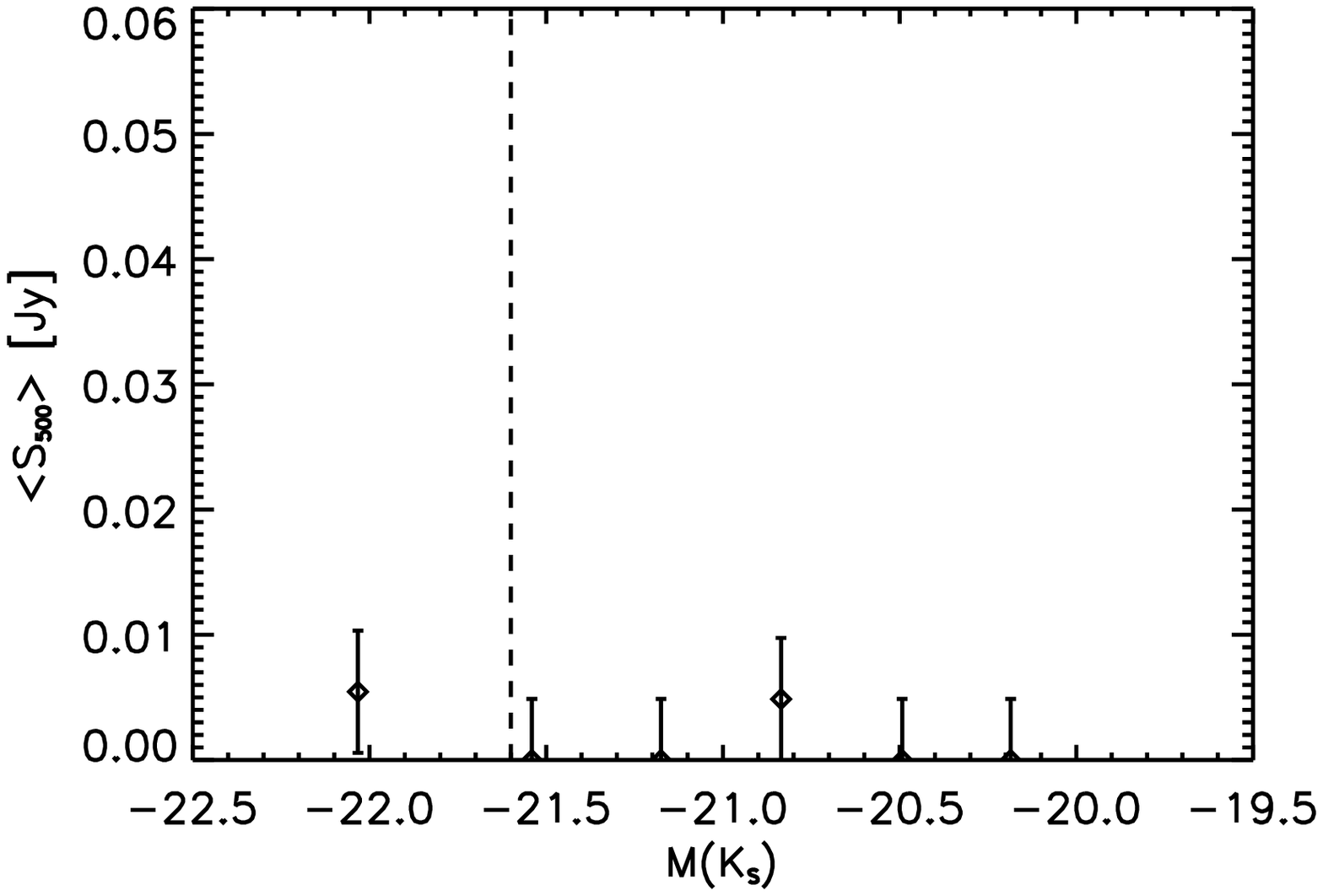}
\caption{Stacked flux density of cluster members in bins of absolute \kmag~magnitude. Left panels include all cluster members; right panels exclude the individual sources. The dashed line marks the position of the characteristic magnitude \kstar. Top row: stacking of 250\,\micron~map. Middle row: stacking of 350\,\micron~map. Bottom row: stacking of 500\,\micron~map.}
\label{stackmag}
\end{figure*}

The result of stacking on magnitude is shown in Figure \ref{stackmag}. A noticeable peak around magnitudes fainter than \kstar~is immediately evident, with the largest part of the BLAST 250-\micron~emission originating from galaxies with \kmag~absolute magnitudes brighter than --21, i.e. around \kstar+1. The stacked flux then decreases beyond \kstar, i.e. towards the giant ellipticals. Given the robust correlation between \kmag~magnitude and stellar mass, we identify these systems as having stellar masses of order of a few times $10^9$ \msun. Removing the BLAST-detected sources shows that a larger number of galaxies contributes to this signal. Stacking the 350\,\micron~map shows a less strong signal, although the transition across \kstar~is still noticeable. Removal of the BLAST counterparts does not change the plot significantly. At 500\,\micron, the mean flux density is consistent with zero at every magnitude.

\section{Discussion}
\label{discuss}

The combination of optical and sub-mm photometry with statistical results from stacking analyses provides insight into the star-formation activity of galaxies in A3112.

We statistically detect the mean sub-mm emission from cluster members, which we quantify as 16.6$\pm$2.5, 6.1$\pm$1.9, and 1.5$\pm$1.3 mJy at 250, 350, and 500\,\micron, respectively. These numbers can provide a reference for future, deeper sub-mm observations of other galaxy clusters. The mean SED and SFR derived from the stacking shows that A3112 has a total SFR density of $12.4 \pm 3.4~\rmn{M}_{\odot}~\rmn{yr}^{-1}/(10^{14} \rmn{M}_{\odot})$. This is in good agreement with previous results (\citealt{Geach06}; \citealt{Haines09b}) and suggests that sub-mm data can help in reducing the scatter in measurements of the evolution of SFR vs. redshift for galaxy clusters.

The results from cluster-centric radial stacking show that a large fraction of the total star-formation activity in the cluster is taking place at specific cluster-centric distances. At radii beyond \rtwo, a large amount of signal is detected, preferentially at 250\,\micron. Previous optical studies of star-formation activity in clusters have shown the presence of radial trends in the relative proportions of star-forming and passive galaxies, across mass scales covering the full range from groups to superclusters (e.g. \citealt{Balogh99}; \cite*{Porter07}; \citealt{Porter08}; \citealt{Wilman08}; \citealt{Braglia09}). It is now well established that at distances of order of \rtwo, the SFRs of infalling galaxies can experience a more or less prolonged period of enhanced activity, mainly due to environmental effects. This is also consistent with the statistical analysis of BLAST sources of \citet{Viero09}, who detect a spatial correlation length of 4.9$\pm$0.7 Mpc for 250\,\micron~sources. This loosely corresponds to about 2\rtwo~for A3112, consistent with the distance of the outermost radial bin in our analysis. The detected signal is thus well explained within the general picture of galaxy infall on to clusters.

In addition, at around 0.6\rtwo~a significant enhancement in the star-formation is detected at BLAST wavelengths. This radius roughly corresponds to \rfive~(\citealt{Sanderson03}), i.e. the boundary between the infall region and the virialized cluster core. The recent investigations of galaxy clusters with {\it Spitzer} have shown that the radial trends detected at optical wavelengths in fact extend deeper into the cluster core, both in filamentary structures (\citealt{Fadda08}) and in the infall regions between \rtwo~and \rfive~(\citealt{Bai07}; \citealt{Saintonge08}; \citealt{Haines09a}; \citealt{Haines09b}; \citealt{Tran09}). In all cases, a global decrease of the fraction of mid-IR luminous galaxies is seen towards the innermost regions of clusters, which is usually interpreted in terms of gradual quenching by environmental effects due to interaction with other cluster members and with the diffuse intra-cluster gas. Our observations in the sub-mm confirm this picture also, at wavelengths which sample a region close to the FIR peak of the cold dust which is heated directly by the UV radiation of O and B stars. Although a large fraction of the sub-mm emission detected is clearly due to the cluster members individually detected at sub-mm wavelengths, we see evidence from stacking of a larger population of galaxies whose sub-mm emission is statistically detected at around \rfive. This suggests that on average, cluster galaxies are expected to show an overall enhanced star-formation level at this distance.

The distribution of mean sub-mm emission with respect to galaxy \kmag~magnitude shows that a significant fraction of star-formation is carried by systems with \kmag$<$\kstar, i.e. by intermediate mass galaxies. The mean sub-mm flux density is seen to increase towards more luminous \kstar~magnitudes, until a noticeable  drop at the brightest magnitudes (as expected for passively evolving ellipticals). This agrees with previous results (e.g. \cite*{Pierini03}) and shows that a large part of the star-formation activity is taking place in intermediate systems.

Analysis of individual sources as well as stacking against colour criteria allows us to assess the nature of the cluster members detected by BLAST. Excluding the single galaxy IRAS~03152--4427, which is identified as a LIRG with a relatively high temperature of 23.4~K and a total SFR of 39~M$_{\odot}~\rmn{yr}^{-1}$, both the SEDs of individual sources and the stacked SED show that typical star-forming galaxies in A3112 are not undergoing strong star-formation activity, but instead are identified as normal star-forming galaxies with total SFRs of a few ($<10$) \sfryr, FIR luminosities of order of $10^{10}$ L$_{\odot}$ and temperatures below 20~K. Their total dust mass is of order 10$^8$ M$_{\odot}$, typical of large spirals. Optical classification of these galaxies reveals them to be massive ($M_{\rmn{Star}} \sim \rmn{[5-10]}\times10^9 \rmn{M}_{\odot}$), disc-like galaxies with relatively red colours. The colour-magnitude diagram reveals that they all lie in a narrow colour strip with $(B-R)$ = 1.38~$\pm$~0.08, i.e. at an interface region between the population of star-forming galaxies and the old passively-evolving galaxies in the RS. The population of galaxies with $(B-R)$ colour redder than the RS is revealed to be composed of quiescent galaxies, whose redder colour is probably due to geometrical properties rather than to increased levels of dust.

\section{Conclusions and summary}
\label{summ}

We have been able to provide a description of the star-formation activity of member galaxies in a cluster at sub-mm wavelengths, as observed by the BLAST experiment. A combination of stacking analyses based on cluster-centric distance, galaxy magnitude and colour allowed us to identify correlations between optical properties of cluster galaxies and their FIR emission.

Studying the sub-mm SEDs of cluster members identifies BLAST-detected cluster members mostly as normal star-forming galaxies. Their optical colours and \kmag~magnitude identify them as massive early-type galaxies in an advanced stage of evolution. The BCG is found to have low dust content, its mid-IR emission possibly due to the central AGN.

Together with an expected increase of the SFR in the cluster outskirts, we find a significant increase in the 250\,\micron~flux density at distances around~\rfive, i.e. further inside the cluster with respect to the infall region. The increase peaks to values 3--4 times as large as in the inner regions of the cluster. This confirms previous studies in the mid-IR that show large fractions of star-forming galaxies closer to the dense cores of galaxy clusters. The combined results of our analysis show that the cluster members identified at sub-mm wavelengths can be part of a population of evolved systems on the verge of transition from the population of blue active galaxies to the quenched systems (ellipticals and S0s) dominating the cluster cores, and suggests that environmental effects at distances of order of \rfive~play a role regulating star-formation activity during this transition.

Deeper and more complete studies of galaxy clusters at far-IR and sub-mm wavelengths with {\it Herschel} and SCUBA-2 should provide more complete coverage of the physical processes at work in cluster galaxies. In particular, combining the higher sensitivity and resolution of SPIRE and the spectral coverage offered by parallel observations with PACS and SPIRE will allow for the detection and characterization of cluster galaxies down to smaller luminosities and masses, providing a more complete description of the star-formation activity in galaxy clusters.

\section{Acknowledgments}

F.G.B. thanks Daniele Pierini for helpful discussions and suggestions.

We acknowledge the support of NASA through grant numbers NAG5-12785, NAG5-13301, and NNGO-6GI11G, the NSF Office of Polar Programs, the Canadian Space Agency, the Natural Sciences and Engineering Research Council (NSERC) of Canada, and the UK Science and Technology Facilities Council (STFC). This research has been enabled by the use of WestGrid computing resources.

This research has made use of the NASA/IPAC Extragalactic Database (NED) which is operated by the Jet Propulsion Laboratory, California Institute of Technology, under contract with the National Aeronautics and Space Administration.

This publication makes use of data products from the Two Micron All Sky Survey, which is a joint project of the University of Massachusetts and the Infrared Processing and Analysis Center/California Institute of Technology, funded by the National Aeronautics and Space Administration and the National Science Foundation.

This work is based in part on observations made with the {\it Spitzer} Space Telescope, which is operated by the Jet Propulsion Laboratory, California Institute of Technology under a contract with NASA.

\bibliographystyle{mn2e}
\bibliography{mn-jour,refs}

\appendix

\section{Catalogues of BLAST sources}

Here we present the full catalogue of all BLAST sources detected in the central 0.8 deg$^2$ of the A3112 field at 250, 350, or 500\,\micron, with a significance of at least $3\sigma$. For each source, its position, flux density and errors are provided. Where flux densities are provided in more than one band, the quoted position is the averaged position of the matched sources (cf. Section \ref{blastdata}). An asterisk ($^{*}$) following the BLAST ID means the source is identified as a counterpart of a cluster member.

\newpage

\begin{table*}
\caption{BLAST catalogue of objects in the field of A3112. Flux densities and errors are given in Jy.}
\centering
\begin{tabular}{c c c c c c c c c c}
\hline
 & BLAST ID & R.A. & Dec. & S$_{250}$ & $\delta$S$_{250}$ & S$_{350}$ & $\delta$S$_{350}$ & S$_{500}$ & $\delta$S$_{500}$ \\
\hline
  1 &      BLAST J031700-441605$^{*}$ &   49.250954 &  -44.268158 &       0.665 &       0.029 &       0.245 &       0.022 &       0.079 &       0.016 \\
  2 &      BLAST J031844-441044$^{*}$ &   49.686970 &  -44.179085 &       0.195 &       0.028 &       0.106 &       0.021 &         0.063 &       0.016 \\
  3 &      BLAST J031829-441138$^{*}$ &   49.622036 &  -44.193897 &       0.181 &       0.028 &       0.090 &       0.021 &         0.054 &         0.015 \\
  4 &      BLAST J031746-442706 &   49.443359 &  -44.451836 &       0.188 &       0.030 &         $-$ &         $-$ &         $-$ &         $-$ \\
  5 &      BLAST J031733-440535 &   49.388805 &  -44.093178 &       0.173 &       0.028 &       0.072 &       0.022 &         $-$ &         $-$ \\
  6 &      BLAST J031803-440636 &   49.514664 &  -44.110165 &       0.167 &       0.028 &       0.096 &       0.022 &         $-$ &         $-$ \\
  7 &      BLAST J031754-441546 &   49.478020 &  -44.262951 &       0.159 &       0.027 &         $-$ &         $-$ &         $-$ &         $-$ \\
  8 &      BLAST J031930-441512$^{*}$ &   49.877411 &  -44.253513 &       0.195 &       0.034 &         $-$ &         $-$ &         $-$ &         $-$ \\
  9 &      BLAST J031636-441359 &   49.153713 &  -44.233124 &       0.176 &       0.032 &       0.189 &       0.025 &         $-$ &         $-$ \\
 10 &      BLAST J031913-441933 &   49.804504 &  -44.325981 &       0.160 &       0.030 &         $-$ &         $-$ &         $-$ &         $-$ \\
 11 &      BLAST J031744-442215 &   49.434784 &  -44.370972 &       0.150 &       0.028 &       0.064 &       0.021 &         $-$ &         $-$ \\
 12 &      BLAST J031909-442124 &   49.788578 &  -44.356796 &       0.164 &       0.031 &         $-$ &         $-$ &         $-$ &         $-$ \\
 13 &      BLAST J031724-442430 &   49.350231 &  -44.408504 &       0.155 &       0.030 &       0.085 &       0.023 &       0.081 &       0.016 \\
 14 &      BLAST J031807-440749 &   49.531059 &  -44.130466 &       0.147 &       0.029 &         $-$ &         $-$ &       0.073 &       0.016 \\
 15 &      BLAST J031808-440826 &   49.535786 &  -44.140617 &       0.134 &       0.028 &         $-$ &         $-$ &         $-$ &         $-$ \\
 16 &      BLAST J031748-442626 &   49.451027 &  -44.440712 &       0.135 &       0.029 &         $-$ &         $-$ &         $-$ &         $-$ \\
 17 &      BLAST J031832-440525 &   49.636398 &  -44.090477 &       0.143 &       0.031 &         $-$ &         $-$ &         $-$ &         $-$ \\
 18 &      BLAST J031622-441434 &   49.094093 &  -44.242931 &       0.153 &       0.034 &         $-$ &         $-$ &         $-$ &         $-$ \\
 19 &      BLAST J031738-441516 &   49.409340 &  -44.254459 &       0.123 &       0.027 &       0.089 &       0.021 &         $-$ &         $-$ \\
 20 &      BLAST J031910-441722 &   49.793259 &  -44.289524 &       0.130 &       0.029 &       0.081 &       0.021 &         $-$ &         $-$ \\
 21 &      BLAST J031848-442637 &   49.703758 &  -44.443665 &       0.138 &       0.032 &         $-$ &         $-$ &       0.056 &       0.016 \\
 22 &      BLAST J031759-441635 &   49.497219 &  -44.276615 &       0.116 &       0.027 &         $-$ &         $-$ &         $-$ &         $-$ \\
 23 &      BLAST J031809-442005 &   49.540203 &  -44.334846 &       0.117 &       0.027 &         $-$ &         $-$ &         $-$ &         $-$ \\
 24 &      BLAST J031725-442418 &   49.354214 &  -44.405098 &       0.124 &       0.029 &       0.085 &       0.023 &       0.081 &       0.016 \\
 25 &      BLAST J031931-441042 &   49.880802 &  -44.178509 &       0.151 &       0.035 &         $-$ &         $-$ &         $-$ &         $-$ \\
 26 &      BLAST J031706-440045 &   49.277191 &  -44.012756 &       0.156 &       0.037 &         $-$ &         $-$ &         $-$ &         $-$ \\
 27 &      BLAST J031839-441435 &   49.663799 &  -44.243259 &       0.113 &       0.027 &         $-$ &         $-$ &         $-$ &         $-$ \\
 28 &      BLAST J031836-441845 &   49.652893 &  -44.312534 &       0.116 &       0.028 &         $-$ &         $-$ &         $-$ &         $-$ \\
 29 &      BLAST J031721-440854$^{*}$ &   49.338924 &  -44.148361 &       0.114 &       0.028 &       0.113 &       0.021 &       0.063 &       0.015 \\
 30 &      BLAST J031829-443201 &   49.624168 &  -44.533665 &       0.142 &       0.037 &         $-$ &         $-$ &       0.080 &       0.020 \\
 31 &      BLAST J031617-442055 &   49.074127 &  -44.348713 &       0.139 &       0.036 &         $-$ &         $-$ &         $-$ &         $-$ \\
 32 &      BLAST J031736-441816 &   49.400284 &  -44.304531 &       0.106 &       0.028 &         $-$ &         $-$ &       0.053 &       0.015 \\
 33 &      BLAST J031603-440944 &   49.013168 &  -44.162308 &       0.144 &       0.038 &         $-$ &         $-$ &         $-$ &         $-$ \\
 34 &      BLAST J031811-440653 &   49.549908 &  -44.114780 &       0.107 &       0.029 &       0.077 &       0.022 &         $-$ &         $-$ \\
 35 &      BLAST J031809-441006 &   49.539707 &  -44.168358 &       0.102 &       0.028 &         $-$ &         $-$ &         $-$ &         $-$ \\
 36 &      BLAST J031727-441253 &   49.365360 &  -44.214848 &       0.101 &       0.027 &       0.081 &       0.021 &       0.051 &       0.015 \\
 37 &      BLAST J031701-442040 &   49.256489 &  -44.344566 &       0.110 &       0.030 &       0.074 &       0.022 &         $-$ &         $-$ \\
 38 &      BLAST J031612-440434 &   49.052586 &  -44.076244 &       0.144 &       0.039 &         $-$ &         $-$ &         $-$ &         $-$ \\
 39 &      BLAST J031748-441006 &   49.450603 &  -44.168423 &       0.101 &       0.028 &         $-$ &         $-$ &         $-$ &         $-$ \\
 40 &      BLAST J031718-441249 &   49.325611 &  -44.213844 &       0.103 &       0.028 &         $-$ &         $-$ &       0.050 &       0.016 \\
 41 &      BLAST J031838-440315 &   49.659569 &  -44.054287 &       0.129 &       0.036 &         $-$ &         $-$ &         $-$ &         $-$ \\
 42 &      BLAST J031800-441746 &   49.501591 &  -44.296204 &       0.100 &       0.028 &         $-$ &         $-$ &         $-$ &         $-$ \\
 43 &      BLAST J031813-441916 &   49.555611 &  -44.321163 &       0.099 &       0.027 &         $-$ &         $-$ &         $-$ &         $-$ \\
 44 &      BLAST J031624-440904 &   49.101841 &  -44.151276 &       0.119 &       0.033 &         $-$ &         $-$ &         $-$ &         $-$ \\
 45 &      BLAST J031638-442455 &   49.159283 &  -44.415443 &       0.126 &       0.035 &         $-$ &         $-$ &         $-$ &         $-$ \\
 46 &      BLAST J031646-442005 &   49.194519 &  -44.334904 &       0.113 &       0.031 &         $-$ &         $-$ &         $-$ &         $-$ \\
 47 &      BLAST J031828-440756 &   49.617111 &  -44.132259 &       0.104 &       0.029 &         $-$ &         $-$ &         $-$ &         $-$ \\
 48 &      BLAST J031647-442035 &   49.198395 &  -44.343300 &       0.109 &       0.031 &         $-$ &         $-$ &         $-$ &         $-$ \\
 49 &      BLAST J031843-441524 &   49.683205 &  -44.256943 &       0.094 &       0.027 &         $-$ &         $-$ &         $-$ &         $-$ \\
 50 &      BLAST J031923-441033 &   49.849792 &  -44.175926 &       0.118 &       0.034 &         $-$ &         $-$ &         $-$ &         $-$ \\
 51 &      BLAST J031820-441704 &   49.584362 &  -44.284637 &       0.095 &       0.027 &       0.065 &       0.020 &       0.062 &       0.015 \\
 52 &      BLAST J031741-442653 &   49.422668 &  -44.448124 &       0.103 &       0.030 &       0.072 &       0.023 &         $-$ &         $-$ \\
 53 &      BLAST J031651-442635 &   49.213417 &  -44.443069 &       0.116 &       0.034 &         $-$ &         $-$ &         $-$ &         $-$ \\
 54 &      BLAST J031749-443116 &   49.454605 &  -44.521275 &       0.130 &       0.038 &         $-$ &         $-$ &         $-$ &         $-$ \\
 55 &      BLAST J031747-442949 &   49.448418 &  -44.497074 &       0.116 &       0.034 &         $-$ &         $-$ &       0.068 &       0.019 \\
 56 &      BLAST J031729-441852 &   49.374840 &  -44.314663 &       0.095 &       0.028 &         $-$ &         $-$ &       0.048 &       0.015 \\
 57 &      BLAST J031846-443014 &   49.692703 &  -44.503994 &       0.116 &       0.034 &         $-$ &         $-$ &         $-$ &         $-$ \\
 58 &      BLAST J031714-442625 &   49.310863 &  -44.440376 &       0.103 &       0.031 &         $-$ &         $-$ &         $-$ &         $-$ \\
 59 &      BLAST J031822-443140 &   49.594753 &  -44.527863 &       0.121 &       0.036 &       0.100 &       0.028 &       0.082 &       0.021 \\
 60 &      BLAST J031957-441030 &   49.990860 &  -44.175240 &       0.151 &       0.046 &       0.109 &       0.031 &         $-$ &         $-$ \\
\hline
\end{tabular}
\end{table*}
 
 \newpage
 
\begin{table*}
\contcaption{}
\centering
\begin{tabular}{c c c c c c c c c c}
\hline
 & BLAST ID & R.A. & Dec. & S$_{250}$ & $\delta$S$_{250}$ & S$_{350}$ & $\delta$S$_{350}$ & S$_{500}$ & $\delta$S$_{500}$ \\
\hline
 61 &      BLAST J031928-441408 &   49.869518 &  -44.235584 &       0.109 &       0.033 &         $-$ &         $-$ &       0.066 &       0.018 \\
 62 &      BLAST J031956-442214 &   49.986340 &  -44.370731 &       0.139 &       0.042 &       0.116 &       0.031 &         $-$ &         $-$ \\
 63 &      BLAST J031620-441125 &   49.086700 &  -44.190289 &       0.113 &       0.034 &         $-$ &         $-$ &         $-$ &         $-$ \\
 64 &      BLAST J031833-443016 &   49.638016 &  -44.504459 &       0.108 &       0.033 &         $-$ &         $-$ &         $-$ &         $-$ \\
 65 &      BLAST J031641-441235 &   49.173447 &  -44.209835 &       0.098 &       0.030 &       0.081 &       0.024 &         $-$ &         $-$ \\
 66 &      BLAST J031648-441226 &   49.202332 &  -44.207226 &       0.096 &       0.030 &         $-$ &         $-$ &         $-$ &         $-$ \\
 67 &      BLAST J031929-441233 &   49.872932 &  -44.209209 &       0.107 &       0.033 &         $-$ &         $-$ &         $-$ &         $-$ \\
 68 &      BLAST J032001-442453 &   50.005367 &  -44.414837 &       0.144 &       0.045 &       0.124 &       0.034 &         $-$ &         $-$ \\
 69 &      BLAST J031815-441346 &   49.563393 &  -44.229546 &       0.087 &       0.027 &         $-$ &         $-$ &         $-$ &         $-$ \\
 70 &      BLAST J031741-442426 &   49.423683 &  -44.407402 &       0.091 &       0.028 &         $-$ &         $-$ &         $-$ &         $-$ \\
 71 &      BLAST J031617-442326 &   49.073780 &  -44.390808 &       0.122 &       0.038 &       0.095 &       0.031 &         $-$ &         $-$ \\
 72 &      BLAST J031656-442546 &   49.236767 &  -44.429703 &       0.102 &       0.032 &         $-$ &         $-$ &         $-$ &         $-$ \\
 73 &      BLAST J031728-443216 &   49.369152 &  -44.537811 &       0.145 &       0.046 &         $-$ &         $-$ &         $-$ &         $-$ \\
 74 &      BLAST J031726-442245 &   49.361435 &  -44.379398 &       0.089 &       0.028 &         $-$ &         $-$ &         $-$ &         $-$ \\
 75 &      BLAST J031910-441755 &   49.794270 &  -44.298737 &       0.093 &       0.030 &       0.081 &       0.021 &         $-$ &         $-$ \\
 76 &      BLAST J031727-443016 &   49.365112 &  -44.504692 &       0.112 &       0.036 &         $-$ &         $-$ &         $-$ &         $-$ \\
 77 &      BLAST J031810-442824 &   49.545727 &  -44.473553 &       0.097 &       0.031 &       0.100 &       0.024 &       0.071 &       0.017 \\
 78 &      BLAST J031734-435956 &   49.392559 &  -43.998978 &       0.112 &       0.036 &         $-$ &         $-$ &         $-$ &         $-$ \\
 79 &      BLAST J031646-440613 &   49.193642 &  -44.103729 &       0.097 &       0.031 &         $-$ &         $-$ &       0.064 &       0.017 \\
 80 &      BLAST J031644-440134 &   49.184139 &  -44.026257 &       0.117 &       0.038 &         $-$ &         $-$ &         $-$ &         $-$ \\
 81 &      BLAST J031811-441106 &   49.547829 &  -44.185001 &       0.084 &       0.027 &         $-$ &         $-$ &         $-$ &         $-$ \\
 82 &      BLAST J031821-440246 &   49.589890 &  -44.046158 &       0.106 &       0.035 &         $-$ &         $-$ &         $-$ &         $-$ \\
 83 &      BLAST J031723-442146 &   49.345963 &  -44.362816 &       0.088 &       0.029 &         $-$ &         $-$ &         $-$ &         $-$ \\
 84 &      BLAST J031706-442716 &   49.275578 &  -44.454510 &       0.097 &       0.032 &         $-$ &         $-$ &         $-$ &         $-$ \\
 85 &      BLAST J031852-442425 &   49.719086 &  -44.407166 &       0.091 &       0.030 &         $-$ &         $-$ &         $-$ &         $-$ \\
 86 &      BLAST J031911-442903 &   49.797077 &  -44.484322 &       0.112 &       0.037 &         $-$ &         $-$ &         $-$ &         $-$ \\
 87 &      BLAST J031758-441552 &   49.493469 &  -44.264530 &         $-$ &         $-$ &       0.151 &       0.021 &       0.120 &       0.015 \\
 88 &      BLAST J031628-441804 &   49.116924 &  -44.301365 &         $-$ &         $-$ &       0.126 &       0.025 &         $-$ &         $-$ \\
 89 &      BLAST J031855-442636 &   49.729778 &  -44.443413 &         $-$ &         $-$ &       0.116 &       0.024 &       0.063 &       0.017 \\
 90 &      BLAST J031749-442706 &   49.454880 &  -44.451839 &         $-$ &         $-$ &       0.100 &       0.022 &         $-$ &         $-$ \\
 91 &      BLAST J031810-442136 &   49.543903 &  -44.360073 &         $-$ &         $-$ &       0.093 &       0.021 &         $-$ &         $-$ \\
 92 &      BLAST J031847-442721 &   49.697227 &  -44.455936 &         $-$ &         $-$ &       0.106 &       0.024 &       0.056 &       0.016 \\
 93 &      BLAST J031934-442819 &   49.893593 &  -44.472050 &         $-$ &         $-$ &       0.138 &       0.031 &       0.102 &       0.022 \\
 94 &      BLAST J031857-442535 &   49.738525 &  -44.426403 &         $-$ &         $-$ &       0.105 &       0.024 &         $-$ &         $-$ \\
 95 &      BLAST J031932-442942 &   49.887066 &  -44.495174 &         $-$ &         $-$ &       0.134 &       0.032 &         $-$ &         $-$ \\
 96 &      BLAST J031907-440404 &   49.783142 &  -44.067883 &         $-$ &         $-$ &       0.127 &       0.031 &         $-$ &         $-$ \\
 97 &      BLAST J031949-442051 &   49.955830 &  -44.347607 &         $-$ &         $-$ &       0.117 &       0.029 &         $-$ &         $-$ \\
 98 &      BLAST J031850-443115 &   49.711781 &  -44.520863 &         $-$ &         $-$ &       0.113 &       0.028 &         $-$ &         $-$ \\
 99 &      BLAST J031749-440636 &   49.454380 &  -44.110088 &         $-$ &         $-$ &       0.084 &       0.021 &         $-$ &         $-$ \\
100 &      BLAST J031952-442101 &   49.967468 &  -44.350323 &         $-$ &         $-$ &       0.117 &       0.030 &         $-$ &         $-$ \\
101 &      BLAST J031622-442544 &   49.093040 &  -44.429081 &         $-$ &         $-$ &       0.120 &       0.032 &         $-$ &         $-$ \\
102 &      BLAST J031745-440325 &   49.439098 &  -44.057167 &         $-$ &         $-$ &       0.084 &       0.023 &       0.067 &       0.017 \\
103 &      BLAST J031900-441744 &   49.753716 &  -44.295624 &         $-$ &         $-$ &       0.076 &       0.021 &         $-$ &         $-$ \\
104 &      BLAST J031616-440504 &   49.067604 &  -44.084553 &         $-$ &         $-$ &       0.113 &       0.031 &         $-$ &         $-$ \\
105 &      BLAST J031753-442706 &   49.474316 &  -44.451923 &         $-$ &         $-$ &       0.081 &       0.022 &         $-$ &         $-$ \\
106 &      BLAST J031715-441746 &   49.315022 &  -44.296326 &         $-$ &         $-$ &       0.081 &       0.022 &         $-$ &         $-$ \\
107 &      BLAST J031814-442046 &   49.559368 &  -44.346241 &         $-$ &         $-$ &       0.074 &       0.020 &         $-$ &         $-$ \\
108 &      BLAST J031716-442036 &   49.318947 &  -44.343410 &         $-$ &         $-$ &       0.076 &       0.021 &         $-$ &         $-$ \\
109 &      BLAST J031602-442153 &   49.011986 &  -44.364788 &         $-$ &         $-$ &       0.128 &       0.036 &         $-$ &         $-$ \\
110 &      BLAST J031601-441624 &   49.004501 &  -44.273369 &         $-$ &         $-$ &       0.107 &       0.031 &         $-$ &         $-$ \\
111 &      BLAST J031908-440524 &   49.787071 &  -44.090057 &         $-$ &         $-$ &       0.101 &       0.029 &         $-$ &         $-$ \\
112 &      BLAST J031703-441935 &   49.264328 &  -44.326649 &         $-$ &         $-$ &       0.076 &       0.022 &         $-$ &         $-$ \\
113 &      BLAST J031857-442314 &   49.738480 &  -44.387379 &         $-$ &         $-$ &       0.076 &       0.023 &         $-$ &         $-$ \\
114 &      BLAST J031702-442235 &   49.260326 &  -44.376656 &         $-$ &         $-$ &       0.076 &       0.023 &         $-$ &         $-$ \\
115 &      BLAST J031617-440954 &   49.074883 &  -44.165115 &         $-$ &         $-$ &       0.094 &       0.028 &         $-$ &         $-$ \\
116 &      BLAST J031710-441935 &   49.295662 &  -44.326645 &         $-$ &         $-$ &       0.072 &       0.022 &         $-$ &         $-$ \\
117 &      BLAST J031609-441544 &   49.039661 &  -44.262333 &         $-$ &         $-$ &       0.093 &       0.028 &         $-$ &         $-$ \\
118 &      BLAST J031817-442908 &   49.573868 &  -44.485615 &         $-$ &         $-$ &       0.078 &       0.024 &       0.051 &       0.017 \\
119 &      BLAST J031723-442639 &   49.349129 &  -44.444225 &         $-$ &         $-$ &       0.075 &       0.023 &       0.107 &       0.017 \\
120 &      BLAST J031812-443015 &   49.552429 &  -44.504337 &         $-$ &         $-$ &       0.083 &       0.026 &         $-$ &         $-$ \\
\hline
\end{tabular}
\end{table*}
 
 \newpage
 
\begin{table*}
\contcaption{}
\centering
\begin{tabular}{c c c c c c c c c c}
\hline
 & BLAST ID & R.A. & Dec. & S$_{250}$ & $\delta$S$_{250}$ & S$_{350}$ & $\delta$S$_{350}$ & S$_{500}$ & $\delta$S$_{500}$ \\
\hline
121 &      BLAST J031638-440935 &   49.160873 &  -44.159855 &         $-$ &         $-$ &       0.077 &       0.024 &       0.058 &       0.017 \\
122 &      BLAST J031900-442324 &   49.753998 &  -44.390053 &         $-$ &         $-$ &       0.073 &       0.023 &         $-$ &         $-$ \\
123 &      BLAST J031803-440016 &   49.516430 &  -44.004478 &         $-$ &         $-$ &       0.095 &       0.030 &         $-$ &         $-$ \\
124 &      BLAST J031710-440235 &   49.292267 &  -44.043304 &         $-$ &         $-$ &       0.082 &       0.026 &         $-$ &         $-$ \\
125 &      BLAST J031724-442630 &   49.350056 &  -44.441772 &         $-$ &         $-$ &       0.072 &       0.023 &       0.107 &       0.017 \\
126 &      BLAST J031628-441405 &   49.117367 &  -44.234879 &         $-$ &         $-$ &       0.080 &       0.026 &         $-$ &         $-$ \\
127 &      BLAST J031715-440546 &   49.315403 &  -44.096146 &         $-$ &         $-$ &       0.069 &       0.022 &         $-$ &         $-$ \\
128 &      BLAST J031658-440235 &   49.241894 &  -44.043251 &         $-$ &         $-$ &       0.084 &       0.027 &         $-$ &         $-$ \\
129 &      BLAST J031709-441846 &   49.287838 &  -44.312916 &         $-$ &         $-$ &       0.069 &       0.022 &         $-$ &         $-$ \\
130 &      BLAST J031745-440126 &   49.439011 &  -44.024052 &         $-$ &         $-$ &       0.080 &       0.026 &         $-$ &         $-$ \\
131 &      BLAST J031842-440105 &   49.678516 &  -44.018135 &         $-$ &         $-$ &       0.100 &       0.033 &         $-$ &         $-$ \\
132 &      BLAST J031638-440755 &   49.160263 &  -44.132133 &         $-$ &         $-$ &       0.076 &       0.025 &         $-$ &         $-$ \\
133 &      BLAST J031808-442435 &   49.536400 &  -44.409927 &         $-$ &         $-$ &       0.065 &       0.021 &         $-$ &         $-$ \\
134 &      BLAST J031808-442206 &   49.536304 &  -44.368374 &         $-$ &         $-$ &         $-$ &         $-$ &       0.076 &       0.015 \\
135 &      BLAST J031930-440953 &   49.876564 &  -44.164803 &         $-$ &         $-$ &         $-$ &         $-$ &       0.098 &       0.020 \\
136 &      BLAST J031803-442546 &   49.513176 &  -44.429489 &         $-$ &         $-$ &         $-$ &         $-$ &       0.074 &       0.016 \\
137 &      BLAST J031842-440915 &   49.679161 &  -44.154236 &         $-$ &         $-$ &         $-$ &         $-$ &       0.066 &       0.015 \\
138 &      BLAST J031849-441745 &   49.706905 &  -44.295883 &         $-$ &         $-$ &         $-$ &         $-$ &       0.062 &       0.015 \\
139 &      BLAST J031622-440804 &   49.094616 &  -44.134716 &         $-$ &         $-$ &         $-$ &         $-$ &       0.081 &       0.019 \\
140 &      BLAST J031837-440215 &   49.655258 &  -44.037655 &         $-$ &         $-$ &         $-$ &         $-$ &       0.083 &       0.020 \\
141 &      BLAST J031937-441922 &   49.905136 &  -44.322906 &         $-$ &         $-$ &         $-$ &         $-$ &       0.075 &       0.018 \\
142 &      BLAST J031843-441005 &   49.679306 &  -44.168129 &         $-$ &         $-$ &         $-$ &         $-$ &       0.063 &       0.016 \\
143 &      BLAST J031801-441936 &   49.505383 &  -44.326775 &         $-$ &         $-$ &         $-$ &         $-$ &       0.058 &       0.015 \\
144 &      BLAST J031827-443115 &   49.614597 &  -44.521065 &         $-$ &         $-$ &         $-$ &         $-$ &       0.080 &       0.020 \\
145 &      BLAST J031707-441306 &   49.280128 &  -44.218430 &         $-$ &         $-$ &         $-$ &         $-$ &       0.061 &       0.016 \\
146 &      BLAST J031629-442535 &   49.124157 &  -44.426395 &         $-$ &         $-$ &         $-$ &         $-$ &       0.075 &       0.021 \\
147 &      BLAST J031949-441701 &   49.955261 &  -44.283859 &         $-$ &         $-$ &         $-$ &         $-$ &       0.067 &       0.020 \\
148 &      BLAST J031711-441005 &   49.299713 &  -44.168232 &         $-$ &         $-$ &         $-$ &         $-$ &       0.051 &       0.015 \\
149 &      BLAST J031621-441814 &   49.089920 &  -44.304119 &         $-$ &         $-$ &         $-$ &         $-$ &       0.062 &       0.019 \\
150 &      BLAST J031907-442434 &   49.781464 &  -44.409515 &         $-$ &         $-$ &         $-$ &         $-$ &       0.056 &       0.018 \\
151 &      BLAST J031702-442935 &   49.260075 &  -44.493210 &         $-$ &         $-$ &         $-$ &         $-$ &       0.069 &       0.022 \\
152 &      BLAST J031926-442422 &   49.859295 &  -44.406387 &         $-$ &         $-$ &         $-$ &         $-$ &       0.057 &       0.019 \\
153 &      BLAST J031906-441314 &   49.776176 &  -44.220661 &         $-$ &         $-$ &         $-$ &         $-$ &       0.048 &       0.016 \\
\hline
\end{tabular}
\end{table*}

\label{lastpage}

\end{document}